\documentclass[aip,
 amsmath,amssymb,
 reprint,%
]{revtex4-1}
\usepackage[thinc]{esdiff}
\usepackage{graphicx}
\usepackage{dcolumn}
\usepackage{bm}
\usepackage[colorlinks=true, allcolors=blue]{hyperref}
\usepackage[utf8]{inputenc}
\usepackage[T1]{fontenc}
\usepackage{mathptmx}
\usepackage{etoolbox}
\usepackage{appendix}
\makeatletter
\def\@email#1#2{%
 \endgroup
 \patchcmd{\titleblock@produce}
  {\frontmatter@RRAPformat}
  {\frontmatter@RRAPformat{\produce@RRAP{*#1\href{mailto:#2}{#2}}}\frontmatter@RRAPformat}
  {}{}
}%
\makeatother

\begin{document}

\preprint{AIP/123-QED}

\title{Theory of Transverse Mode Instability in Fiber Amplifiers \\
with Multimode Excitations}
\author{Kabish Wisal}
\affiliation{Dept. of Physics, Yale University, New Haven, CT 06520, USA}
\email{kabish.wisal@yale.edu}
\author{Chun-Wei Chen}%
\author{Hui Cao}
\author{A. Douglas Stone}
\affiliation{Dept. of Applied Physics, Yale University, New Haven, CT 06520, USA}



\date{\today}



\begin{abstract}
Transverse Mode Instability (TMI) which results from dynamic nonlinear thermo-optical scattering is the primary limitation to power scaling in high-power fiber lasers and amplifiers. It has been proposed that TMI can be suppressed by exciting multiple modes in a highly multimode fiber. We derive a semi-analytic frequency-domain theory of the threshold for the onset of TMI in narrowband fiber amplifiers under arbitrary multimode input excitation for general fiber geometries under. Our detailed model includes the effect of gain saturation, pump depletion and mode dependent gain. We show that TMI results from exponential growth of noise in all the modes at downshifted frequencies due to the thermo-optical coupling. The noise growth rate in each mode is given by the sum of signal powers in various modes weighted by pairwise thermo-optical coupling coefficients. We calculate thermo-optical coupling coefficients for all $\sim$$10^4$ pairs of modes in a standard circular multimode fiber and show that modes with large transverse spatial frequency mismatch are weakly coupled resulting in a banded coupling matrix. This short-range behavior is due to the diffusive nature of the heat propagation which mediates the coupling and leads to a lower noise growth rate upon multimode excitation compared to single mode, resulting in significant TMI suppression. We find that the TMI threshold scales linearly with the number of modes that are excited asymptotically, leading to roughly an order of magnitude increase in the TMI threshold in a 82-mode fiber amplifier.
\end{abstract}

\maketitle

\section{Introduction}

Fiber lasers based on multi-stage fiber amplifiers (FA) provide an efficient and compact platform to generate ultra-high laser power\cite{even2002high,jeong2004ytterbium,nilsson2011high,Cesar2013high,Zervas2013high,dong2016fiber}. This has enabled potential applications in a wide range of technologies such as laser welding\cite{Welding2018}, gravitational wave detection\cite{buikema2019narrow}, and directed energy\cite{Defense2019}. To realize fully this potential, further power scaling is needed in the FAs\cite{dawson2008analysis}, which has been limited primarily by nonlinear effects such as transverse mode instability (TMI)\cite{Cesar2020transverse,eidam2011experimental,smith2011mode,Ward2012origin} and stimulated Brillouin scattering (SBS)\cite{boyd2020nonlinear,kobyakov2010stimulated,wolff2021brillouin}. TMI is a dynamic transfer of power between the transverse modes of the fiber caused by thermo-optical scattering and/or inversion fluctuations\cite{Hansen2013theoretical,dong2013stimulated,Ward2013modeling,ward2016theory,zervas2019transverse}. During the optical amplification process, heat is generated due to the quantum defect, in an amount proportional to the local optical intensity, creating local temperature fluctuations. The local temperature variations then create refractive index variations, causing significant scattering between the modes. Consequently, when the FA is operated above a certain output power, defined as the TMI threshold, a significant degradation of beam quality occurs\cite{jauregui2011impact, jauregui2012physical, otto2012temporal, naderi2013investigations, kuznetsov2014low, kong2016direct, stihler2018modal,  stihler2020intensity, menyuk2021accurate, chu2021increasing, ren2022experimental}, rendering the output unsuitable for many applications. As a result, suppressing TMI or equivalently raising the TMI threshold has been one of the most important and technologically relevant scientific goals in the high-power laser community\cite{Zervas2013high,Cesar2020transverse,Cesar2013high}. 

Significant efforts have been undertaken to mitigate TMI, such as dynamic seed modulation\cite{Otto2013controlling}, synthesizing materials with low thermo-optic coefficient\cite{Dragic2021kilowatt,dong2023power}, modulating the pump beam\cite{tao2015mitigating,jauregui2018pump}, increasing the optical loss of Higher Order Modes (HOMs)\cite{stutzki2011high,Eidam2011preferential,jauregui2013passive,Robin2014modal,haarlammert2015optimizing,tao2016suppressing}, and utilizing multicore fibers\cite{Otto2014scaling,klenke2022high, abedin2012cladding}. Although, most of these efforts have had some success, they suffer from one or more drawbacks, such as fluctuating output power in the case of seed or pump modulation, difficulty with mass-manufacturing custom fibers, and increased guidance of HOMs due to fiber heating, rendering the HOM suppression unfeasible. As such, efficient TMI suppression remains a highly active area of research.

A common feature in all of the previous approaches is to excite the fundamental mode (FM) of the fiber as much as possible, hence reducing the excitation and amplification of HOMs\cite{stutzki2011high,Eidam2011preferential,hansen2011thermo,jauregui2013passive,Robin2014modal,haarlammert2015optimizing,tao2016suppressing,Otto2014scaling,klenke2022high,tao2015mitigating,jauregui2018pump,Dragic2021kilowatt,dong2023power,Otto2013controlling}. This is motivated by the widespread impression that the speckled internal field generated by multimode excitation will necessarily create a poor beam quality at the  output\cite{zhan2009degradation,chu2019experimental}.  However, recent progress in wavefront shaping, enabled by spatial light modulators (SLM), has demonstrated that multimode excitation with a narrowband seed laser does {\em not} inherently lower the beam quality as long as the light remains spatially coherent\cite{Florentin2017shaping,geng2021high,Cizmar2018three}. In fact, by using an SLM to wavefront-shape the input light to a fiber, it is possible to obtain a diffraction-limited spot after coherent multimode propagation in both passive~\cite{gomes2022near} and active fibers~\cite{Florentin2017shaping}, which can be easily collimated using a lens. This enables a fundamentally novel approach to control nonlinear effects in fibers by utilizing selective multimode excitation\cite{teugin2020controlling,wright2015controllable, krupa2019multimode, qiu2023spatiotemporal}. The viability of such an approach has recently been demonstrated for other nonlinear effects in fibers such as SBS\cite{wisal2022generalized,wisal2023theory,chen2023mitigating} and stimulated Raman scattering (SRS)\cite{tzang2018adaptive}.

The current authors recently showed, using numerical simulations along with some initial theoretical results, that sending power in multiple fiber modes robustly raises the TMI threshold\cite{chen2023suppressing}. To investigate this approach thoroughly and in detail, we have developed a theory of TMI for arbitrary multimode input excitations and general fiber geometries, building on significant prior theoretical efforts to model TMI, which have been successful in capturing much of the key physics\cite{Ward2012origin,smith2013steady,Hansen2013theoretical,dong2013stimulated,Smith2013increasing,naderi2013investigations,hansen2014impact,tao2015study,jauregui2015simplified,zervas2017transverse,tao2018comprehensive,zervas2019transverse,menyuk2021accurate,dong2022accurate}. It has been shown that TMI can be modelled as exponential growth of noise in the HOMs due to the thermo-optical scattering of the signal. However, as noted, almost all of the previous efforts have focused on the case when only the FM of the fiber is excited with presence of noise in a few HOMs. None of these works attempted to model highly multimode excitations, and their formalisms were not suited for such explorations. As mentioned, some key theoretical results for TMI threshold upon multimode excitation were presented in in Ref.~\cite{chen2023suppressing} by the authors of this paper, including results demonstrating linear scaling of the TMI threshold with the number of excited modes. This previous work focused on time-domain numerical simulations, and the detailed theoretical formalism for TMI threshold along with the analysis of pump depletion and gain saturation for fibers with arbitrary geometries and input excitations was deferred to the current work.

Here we develop a general formalism that allows efficient calculation of the TMI threshold for arbitrary multimode input excitations and fiber geometries in narrow linewidth fiber amplifiers. We derive coupled amplitude equations starting from coupled optical propagation and heat diffusion equations and solve them to obtain analytic formulas for the TMI gain and thermo-optical coupling coefficient. A key result of our theory is that the thermo-optical coupling is strong only between the optical modes that have similar transverse spatial frequencies. This is a result of the diffusive nature of the heat propagation, which underlies the thermo-optical coupling. This leads to a banded thermo-optical coupling matrix, due to which the effective TMI gain is significantly lowered upon multimode excitation, resulting in robust TMI suppression. We show that the TMI threshold increases linearly with the number of excited modes, leading to more than an order of magnitude higher TMI threshold in standard multimode fibers. We also show that threshold enhancement can be further increased by tailoring the the fiber geometry.

\begin{figure*}
    \centering
    \includegraphics[width=0.95\textwidth]{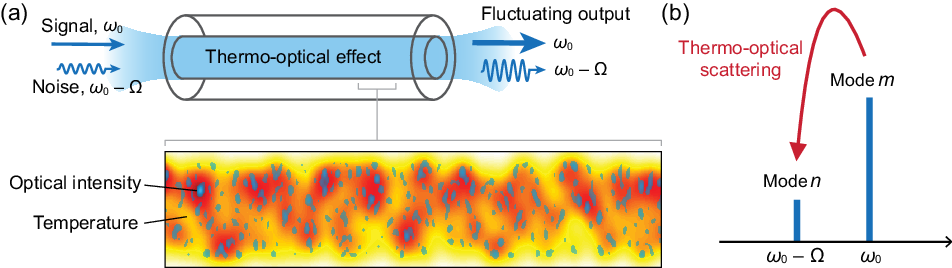}
    \caption{{\bf Schematic of TMI in a multimode fiber amplifier.} At the fiber input, a signal wave is launched in multiple modes at the same frequency, $\omega_0$, with a small amount of noise at the shifted Stokes frequency, $\omega_0-\Omega$. As the signal propagates in the fiber, it undergoes amplification due to the stimulated emission, generating heat in the process and causing temperature fluctuations (shown in red in the inset). These temperature variations result in refractive index variations due to the thermo-optical effect, which cause scattering of power between various modes. (Note:longitudinal and transverse dimensions in lower panel are not to scale). (b) The signal in each mode $m$ at $\omega_0$ transfers power to the stokes shifted frequency in every other mode $n$, which causes the noise to grow exponentially with a growth rate depending on the signal power and pairwise thermo-optical coupling coefficients. When the noise at the output becomes a significant fraction of the signal, the output beam becomes unstable with dynamic fluctuations on a millisecond timescale. The output power at the onset of significant fluctuations (typically set at $1\%$) is defined as TMI threshold.}
    \label{fig:Fig1}
\end{figure*}

We begin by deriving a semi-analytic solution for the optical and heat equations, coupled via the nonlinear polarization and quantum defect heating terms. We expand the optical fields and the temperature fluctuations in terms of eigenmodes of the optical and heat equations respectively. From the driven heat equation, the coefficient of each temperature eigenmode is found in terms of optical amplitudes, leading to coupled amplitude equations for the optical modes, resembling those for a four-wave-mixing process\cite{cronin1984theory}. Next, we solve these coupled amplitude equations for an arbitrary input signal at frequency, $\omega_0$, along with the presence of small amount of noise at stokes shifted frequencies, $\omega_0-\Omega$, in each mode. We show that the noise power in various modes grows exponentially, and can lead to dynamic spatial fluctuations in the output beam profile after sufficiently high noise growth. The growth rate of the noise power in a given mode is equal to the sum of the signal power in each of the other modes, weighted by pairwise thermo-optical coupling coefficients. For any pair of optical modes, the thermo-optical coupling coefficient is proportional to the overlap integrals of the two optical mode profiles with various eigenmodes of the heat equation. The coupling between the optical modes with a large separation in transverse spatial frequency is mediated by temperature modes with high transverse spatial frequency; these thermal modes necessarily have a very short-lifetime, leading to a quite weak coupling between such mode pairs. To study this explicitly, we calculate the thermo-optical coupling coefficients for all $\sim10^4$ mode pairs in a commercially available highly multimode (MM) circular step-index fiber (supporting 82 modes per polarization) and find that the matrix of peak values of pairwise thermo-optical coupling coefficients is a banded, effectively sparse matrix. This sparse/banded nature of the thermo-optical coupling is not affected by the presence of gain saturation or pump depletion. As a result, the noise growth rate goes down linearly with the number of modes excited, if the power is equally (or nearly equally) distributed in all the excited modes. Therefore the TMI threshold is predicted to increase linearly with the number of equally excited modes, which is verified by explicit calculation of the TMI threshold. 

The generality of our approach allows us to do similar calculations for different fiber cross-sections (e.g. square and D-shaped); we find that the linear increase of the threshold is generic and only weakly sensitive to fiber and cladding geometry. Our theory is presented first with a simpler model which neglects the effects of gain saturation and pump depletion, but highlights the universality of the physics due to the mismatch of spatial and temporal scales (thermal vs. optical), which mediates TMI suppression via multimode excitation. In the final section of the paper we present an improved model which does include these important effects (at a tractable computational cost), and confirms that the asymptotic linear scaling of the TMI threshold is preserved with a significant but modest increase in the calculated TMI threshold. Taken in full, our results quantitatively confirm that highly multimode excitations efficiently suppress TMI, opening up a new platform for robust power scaling in high-power fiber laser amplifiers.

\section{Theory and Results}
\subsection{Coupled Amplitude Equations}

TMI is a result of dynamic transfer of power between fiber modes due to the thermo-optical scattering\cite{Cesar2020transverse,Hansen2013theoretical,dong2013stimulated}. A schematic of the TMI process is shown in Fig.~\ref{fig:Fig1}. A signal wave at frequency $\omega_0$ is sent through an active fiber to undergo optical amplification and generate the output. But due to imperfect coupling or other experimental artifacts, there usually exists a small amount of noise at shifted frequencies, $\omega_0-\Omega$, in various modes. The optical signal and noise can interfere to create a spatio-temporally varying optical intensity pattern, which due to the quantum-defect heating that accompanies stimulated emission, results in dynamic temperature variations\cite{jauregui2012physical}. These spatio-temporal thermal fluctuations result in refractive index variations, which can cause significant transfer of power from signal at $\omega_0$ in one mode to the noise at Stokes-shifted frequencies $\omega_0-\Omega$ in other modes. As a result, at high-enough output power, the noise in various modes can become a significant fraction of the output power, leading to fluctuations in the output beam profile. This output power level is defined as the TMI threshold\cite{dong2013stimulated,Hansen2013theoretical} and marks the onset of unstable regime for the output. To derive a formalism to calculate the TMI threshold, we solve the optical wave equation with a temperature dependent nonlinear polarization as a source term\cite{dong2013stimulated}: 
\begin{equation}
    \nabla^2 \vec{E} - \frac{n_0^2}{c^2}\frac{\partial^2\vec{E}}{\partial t^2}=\mu_0\frac{\partial^2 \vec{P}_{\rm NL}}{\partial t^2},
    \label{Eq:OptEqn}
\end{equation}
\noindent
where $\vec{E}$ is the total electric field , $c$ is the speed of light and $n_0$ is the linear refractive index of the fiber. The temperature-dependent polarisation is given by:
\begin{equation}
   \vec{ P}_{\rm NL}=\epsilon_0 \eta \Delta T \vec{E}.
   \label{Eq:PNL}
\end{equation}
\noindent
Here, $\eta$ is the thermo-optic coefficient of the fiber material and $\Delta T$ represent the local temperature fluctuation due to quantum-defect heating. $\Delta T$ satisfies the heat equation with a source term proportional to the local intensity\cite{Hansen2013theoretical}:
\begin{equation}
    [\nabla^2- \frac{\rho C}{\kappa} \frac{\partial}{\partial t}] \Delta T= \frac{Q}{\kappa},
    \label{Eq:heatEq}
\end{equation}
\begin{equation}
    Q(\vec{r},t)=g q_{\rm D} I(\vec{r},t). 
    \label{Eq:heatSrc}
\end{equation}
\noindent
The heat source $Q$ is proportional to the local optical intensity $I$, the linear gain coefficient $g$, and the quantum defect $q_{\rm D}=(\frac{\lambda_{\rm s}}{\lambda_{\rm p}}-1)$, which depends on the difference between the signal wavelength $\lambda_{\rm s}$ and pump wavelength $\lambda_{\rm p}$. The thermal conductivity, the specific heat, and the density of the material are given by $\kappa$, $C$ and $\rho$, respectively. The ratio, ${\kappa}/{\rho C}$ is equal to the diffusion constant $D$. Since the thermo-optical polarization depends on the temperature, which in turn depends on the optical intensity, the optical and heat equations are coupled and nonlinear due to the source terms on the right hand side (RHS) of Eq.~\ref{Eq:OptEqn} and Eq.~\ref{Eq:heatEq}. For a translationally invariant system, such as an optical fiber, both optical and heat equations can be solved formally by expanding the fields in terms of the eigenmodes of the linear operator corresponding to each equation. As usual, we decompose the electric field $\vec{E}$ as a sum of the product of transverse mode profile, a rapidly varying longitudinal phase term, and slowly varying amplitude (SVA) for each fiber mode:
\begin{figure}
    \centering
    \includegraphics[width=0.45\textwidth]{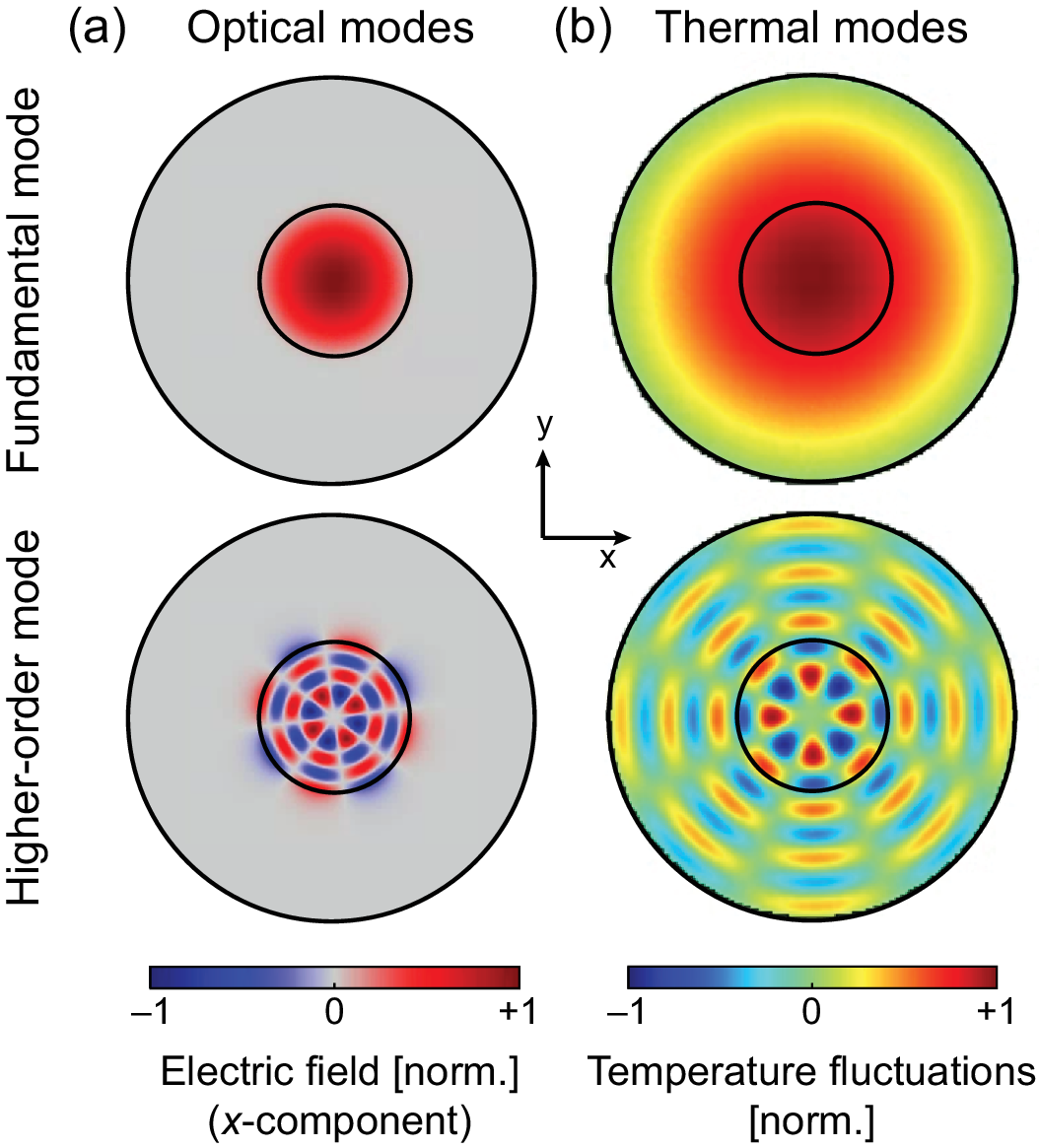}
    \caption{(a) Amplitude profiles of guided optical fundamental mode (FM) and a higher order mode (HOM) for a circular step-index fiber. The optical modes are guided in the core of the fiber. The FM (LP01) has no nodes whereas each HOM denoted as LPuv is characterized by $u$ azimuthal nodes and $v-1$ radial nodes. Here LP44 is shown. (b) Amplitude profiles for thermal modes, which are the spatial eigenmodes of the heat equation with constant temperature at the outer cladding. Each thermal mode fills the entire fiber cross section. Similar to optical modes, FM has no nodes whereas each HOM denoted is characterized by an integer number of radial and azimuthal nodes. Cladding and core sizes are not to scale. }  
    \label{fig:circ_modes}
\end{figure}
\begin{equation}
    \vec{E}(\vec{r},t)=\int_{-\infty}^{\infty}d\Omega \sum_{m}A_m(\Omega,z) \vec{\psi}_m(r_{\bot})e^{i((\omega_0-\Omega)t-\beta_m z)}+{\rm c.c.}
    \label{Eq:Eanstaz}
\end{equation}
\noindent
Here, $A_{m}(\Omega,z)$ is the SVA in mode $m$ at a point $z$ along the fiber axis and for a Stokes frequency shift, $\Omega$ from the central frequency $\omega_0$. The $\Omega=0$ amplitude corresponds to the signal, and $\Omega\neq0$ corresponds to the noise. $\vec{\psi}_m$ and $\beta_m$ denote the transverse mode profile and propagation constant for mode $m$ and can be obtained by solving the fiber modal equation\cite{okamoto2021fundamentals,snyder1978modes}:
\begin{equation}
    \nabla_T^2 \vec{\psi}_m + (\frac{n_0^2 \omega_0^2}{c^2}- \beta_m^2) \vec{\psi}_m=0.
    \label{Eq:Emodal}
\end{equation}
The solutions to Eq.~\ref{Eq:Emodal} correspond to the guided fiber modes, which are confined to the fiber core by total internal reflection. These modes can be calculated using a numerical solver such as COMSOL\cite{comsol} for an arbitrary fiber cross-section geometry. As an example, in Fig.~\ref{fig:circ_modes}a we have shown electric field profiles for the FMs and HOMs for a circular step-index fiber. The width of the core in each case in 40 $\mu \rm m$ and the cladding width is 200 $\mu \rm m$.  The FM have no node in the electric field profile and follow the symmetries of the cross-section geometry. Each HOM can be labelled as linearly polarized (LP) mode and characterized by two indices $(u,v)$ in the standard notation\cite{okamoto2021fundamentals}. The mode ${\rm LP}uv$ has $u$ azimuthal nodes and $(v-1)$ radial nodes, with a radial profile given by the $(u+1)^{\rm th}$ Bessel function of the first kind, with $v-1$ zeros within the fiber core and angular profile given by either a sine or cosine function with $u$ zeros. Figure~\ref{fig:circ_modes}a shows the profiles of $\rm LP01$ (FM) and $\rm LP44$ (HOM) modes. In appendix A, we study fibers with different geometries (D-shaped and Square cross-sections) and the associated optical modes are shown in Fig.~A1a.

Normalizing each mode profile $\psi_m$ to have unit power, the total power in mode $m$ at point $z$ along the fiber axis and at Stokes frequency $\Omega$ is given by ${|A_m(z,\Omega)|}^2$. The heat source term $Q$ in Eq.~\ref{Eq:heatEq} can be obtained in terms of optical modal amplitudes using Eq.~\ref{Eq:heatSrc} and Eq.~\ref{Eq:Eanstaz}. The optical intensity $I$ contains terms corresponding to the interference of various fiber modes,  and thus $Q$ ($\propto I$) is given by a sum of dynamic heat sources oscillating at Stokes frequency $\Omega$ and propagation constant $q_{ij}=\beta_j-\beta_i$ for each mode pair $\{i,j\}$. Thus, we expand $\Delta T$ into a series of temperature profiles corresponding to each heat source term: 
\begin{equation}
    \Delta T(\vec{r},t) = \int_{-\infty}^{\infty}d\Omega \sum_{i,j} T_{ij}(\vec{r},\Omega) e^{i(\Omega t - q_{ij}z)}.   
    \label{Eq:Tansatz}
\end{equation}
Here $T_{ij}(\vec{r},\Omega)$ is the profile of the temperature grating with longitudinal wavevector $q_{ij}$ and temporal frequency $\Omega$. We further decompose each temperature profile $T_{ij}$ in terms of the eigenmodes of the transverse Laplacian $\nabla_T^2$:
\begin{equation}
 T_{ij}(\vec{r},\Omega)=\sum_k a^k_{ij}(z,\Omega) \Tilde{T}^k(\vec{r}_\perp). 
 \label{Eq:Tijansatz}
\end{equation}
\noindent
Here, $a^k_{ij}$ is the SVA for temperature eigenmode $k$ in temperature grating $\{i,j\}$. Temperature eigenmodes of the fiber oscillate in space and relax exponentially in time due to diffusive heat equilibration.  Their decay rate is proportional to the thermal diffusion constant and the wavevector of the fluctuations squared. Hence long wavelength fluctuations dominate the thermal response; as we will see, this leads directly to a nonlinear thermo-optic coupling which is short-range in the momentum {\it difference} between the optical modes. $\Tilde{T}^k$ denotes the transverse profile of temperature eigenmode $k$ which satisfies the following eigenvalue equation:
\begin{equation}
    \nabla_T^2\: \Tilde{T}^k(\vec{r}_\perp)= -\alpha_k^2\: \Tilde{T}^k (\vec{r}_\perp)\:\:,\:\:\:\:
    \Tilde{T}^k|_{\partial \Omega}=0.
    \label{Eq:Tmodal}
\end{equation}

\noindent $\nabla_T^2$ is a self-adjoint operator, thus its eigenmodes $\{\Tilde{T}^k\}$ form a complete and orthogonal basis, with $-\alpha_k^2$ as
the eigenvalue for eigenmode $k$. We impose Dirichlet boundary conditions at the outer surface of the fiber, corresponding to the standard assumption\cite{dong2013stimulated} that outer surface is held at a constant temperature, and all the heat immediately dissipates into a heat bath. The temperature eigenmodes can be solved either analytically or numerically depending on the fiber geometry. We have shown in Fig.~\ref{fig:circ_modes}b the fundamental and higher-order temperature eigenmodes for a fiber with circular cladding, which are calculated using the Coefficient-Form-PDE module in COMSOL\cite{comsol}.
The spatial structure of the temperature eigenmodes are similar to the optical modes since both are eigen-functions of same spatial operator $\nabla_T^2$. An important distinction is that optical modes are localised in the fiber core, but the temperature eigenmodes spread out over the entire fiber cross-section, due to the differing boundary conditions. The fundamental mode has no nodes and follow the symmetries of the cladding geometry. The higher order eigenmodes are characterised by $u$ azimuthal and $v$ radial nodes across the entire fiber cross-section (i.e., core plus cladding), with a radial dependence given by $(u+1)^{\rm th}$ Bessel function of the first kind and angular dependence given by sine or cosine with $u$ zeros. 

To complete the solution for $\Delta T$, we need to determine each coefficient $a^k_{ij}$. This is done by simplifying the LHS of Eq.~\ref{Eq:heatEq} by substituting $\Delta T$ from Eq.~\ref{Eq:Tansatz} and Eq.~\ref{Eq:Tijansatz} and using the eigenmode equation Eq.~\ref{Eq:Tmodal}. The RHS of Eq.~\ref{Eq:heatEq} can be obtained in terms of optical amplitudes by using Eq.~\ref{Eq:heatSrc} and Eq.~\ref{Eq:Eanstaz}. Finally, by exploiting orthogonality, we isolate the desired coefficient $a^k_{ij}$ by multiplying both sides of simplified Eq.~\ref{Eq:heatEq} with $\Tilde{T}^{k*}$ and integrating it across the fiber cross-section leading to:
\begin{equation}
\begin{aligned}
a^{ij}_k(z,\Omega)=\frac{D\langle\vec{\psi}_i^*\cdot\vec{\psi}_j \Tilde{T}_k^* \rangle}{\Gamma_k^{ij}+i\Omega} \frac{gq_{\rm D}}{\kappa} \int_{-\infty}^{\infty} & d\Omega' A_i^*(z,\Omega') \\
& A_j(z,\Omega+\Omega').
\label{Eq:temp_amplitude}
\end{aligned}
\end{equation}
The amplitude of temperature eigenmode $k$ for the heat source term $\{i,j\}$ is proportional to the overlap of the dot product of $\vec{\psi}_i$ and $\vec{\psi}_j$ with the temperature mode profile $\Tilde{T}^k$ integrated over the fiber cross-section, denoted by the angular brackets $\langle.\rangle$. It is also proportional to the convolution of optical amplitudes $A_i$ and $A_j$. We introduce an inverse mode lifetime for mode $k$ given by $\Gamma_k^{ij}={((\beta_i-\beta_j)}^2+ \alpha_k^2 )D$ with units of $[s^{-1}]$. The first term in the inverse mode lifetime corresponds to the longitudinal heat diffusion, and the second term corresponds to the transverse heat diffusion. In typical fibers, the transverse heat diffusion dominates, so the longitudinal heat diffusion can be ignored\cite{Hansen2013theoretical,dong2013stimulated}. In that case, we can define $\Gamma_k^{ij}\approx\Gamma_k = \alpha_k^2D$, which we use in the rest of this paper. However, in fibers where longitudinal diffusion is significant, both contributions need to be included in the inverse mode lifetime. This completes the formal solution for $\Delta T$ in terms  of the optical mode profiles and amplitudes. 

Next, we use the solution for $\Delta T$ to evaluate the source terms in Eq.~\ref{Eq:OptEqn}, which can be used to derive coupled self consistent equations for optical amplitudes. We use the ansatz for $\vec{E}$ from Eq.~\ref{Eq:Eanstaz} to simplify the left hand side (LHS) of Eq.~\ref{Eq:OptEqn}. Finally we use the orthogonality of the optical modes to isolate the equation of each optical amplitude $\{A_m\}$ by multiplying both sides of simplified Eq.~\ref{Eq:OptEqn} with $\vec{\psi}_m^*$ and integrating it over the fiber cross-section:
\begin{equation}
\begin{aligned}
\diff{ A_m(z,\Omega)}{z}  =&\frac{g}{2}A_m + ig \chi_0  \sum_{ijn} \int d\Omega' A_n(z,\Omega-\Omega')\\&  G_{mn ij}(\Omega') \int d\Omega^{''}  A_i^*(z,\Omega^{''})A_j(z,\Omega'+\Omega^{''}) \\&  e^{i(\beta_m-\beta_n+\beta_i-\beta_j)z}.
\end{aligned}
\label{Eq:CAS}
\end{equation}
\noindent
The $z$ derivative of the modal amplitude $A_m$ has contributions from both the linear amplification term with growth rate $g/2$ due to stimulated emission and a nonlinear four-wave-mixing term\cite{cronin1984theory} due to the thermo-optical interaction, which is proportional to  a susceptibility constant $\chi_0$ and the sum of convolutions of three modal amplitudes $A_n$, $A_i^*$, and $A_j$. The exponential term tracks the phase mismatch for each contribution, and the strength of each term is proportional to the overlap of the Green's function of the heat equation with the relevant optical mode profiles:
\begin{equation}
   G_{mn ij}(\Omega) = \sum_{k}\frac{D\langle\vec{\psi}_i^*\cdot\vec{\psi}_j \Tilde{T}_k^* \rangle\langle\Tilde{T}_k \psi_m^*\psi_n \rangle}{\Gamma_k+i\Omega}, \:\:\:\:  \chi_0=\frac{\eta q_{\rm D} k_0}{2 n_c \kappa}.
\end{equation}
\noindent
The sum over $k$ represents the sum over all the temperature eigenmodes. Thus, $G_{mnij}(\Omega)$ denotes the optical modal overlap with the Green's function of the heat equation written in the spectral representation\cite{cole2010heat}. The susceptibility constant $\chi_0$ involves a combination of various material and optical constants involved and has dimensions of $[W^{-1}]$. $G_{mnij}(\Omega)$ has both real and imaginary parts in general. The real part is responsible for the nonlinear phase evolution of $A_m$, while the imaginary part is responsible for dynamic transfer of power between various modes. Clearly, for $\Omega=0$, the imaginary part of $G_{mnij}$ vanishes and thus there is no transfer of power between various modes of the fiber at $\omega_0$. A nonzero Stokes frequency shift is needed for TMI, which is in agreement with the well known result that a moving intensity grating ($\Omega\neq0$) is required for TMI\cite{jauregui2012physical,Hansen2013theoretical,Robin2014modal,Cesar2020transverse}. Typically there is amplitude noise in the fiber, either due to experimental imperfections or due to spontaneous emission, having multiple frequency components. The noise frequency at which imaginary part of Green's function peaks undergoes the highest amplification and is typically a few $~kHz$, leading to dynamic fluctuations in the beam profile on the order of milliseconds\cite{Hansen2013theoretical,jauregui2011impact,Cesar2020transverse}.

For any given set of input amplitudes in each mode, $\{A_m(z=0,\Omega)\}$, Eq.~\ref{Eq:CAS} can be solved to obtain modal amplitudes $\{A_m(z,\Omega)\}$ everywhere, fully determining both $\vec{E}$ and $\Delta T$. The fiber properties are taken into account through $\chi_0$ and $G_{mnij}$  via optical mode profiles $\{\psi_m\}$ and temperature mode profiles $\{\Tilde{T}^k\}$. Equation~\ref{Eq:CAS} is a set of highly coupled nonlinear differential equations, and in general cannot be solved analytically. Numerical solutions are possible by using finite difference methods\cite{sun2023finite} to discretize the derivative operator and evaluating the convolutions by either built-in or custom operations. It is expected that numerically solving Eq.~\ref{Eq:CAS} would be more computationally efficient than solving the original coupled optical and heat equations. This is because Eq.~\ref{Eq:CAS} is a set of 1-D equations, where transverse degrees of freedom are accounted through the fiber modes, which exploit the longitudinal translation invariance and only need to be solved once for a given fiber. However, despite this simplification, when a large number of modes are present, the computational complexity can quickly become very high. Additionally, for studying TMI suppression using multimode excitation, the input modal content needs to be parametrically tuned, requiring a fast and efficient solution. Fortunately, an approximate solution to the coupled amplitude equations can be obtained that captures the essential features of the onset of TMI and can be used to calculate the TMI threshold, which is derived in the next section.

\subsection{Phase-Matched Noise Growth}

In this section, we derive an approximate solution to the coupled amplitude equations to study noise growth due to thermo-optical scattering when signal power is launched into multiple fiber modes. We utilize two key approximations: (1) we retain only the phase-matched terms\cite{Hansen2013theoretical,dong2013stimulated,dong2022accurate,menyuk2021accurate}, since phase-mismatched terms become insignificant over long enough length scales, and (2) we assume that the noise power is significantly lower than the signal power, and the change in the signal due to the noise growth is ignored\cite{zervas2017transverse,Hansen2013theoretical,dong2013stimulated}. This is typically valid below the TMI threshold, which marks the onset of significant beam fluctuations, and can be used to calculate the threshold. We assume that at the fiber input, the signal (or, seed) is injected at frequency $\omega_0$ in various modes with complex amplitudes $\{A^{\rm s}_m(z=0)\}$. In addition, there is noise present at Stokes shifted frequencies $\omega_0-\Omega$ in all the fiber modes, denoted by complex amplitudes $\{B_m(z=0,\Omega)\}$. The total input amplitude in mode $m$ can be written as:
\begin{equation}
    A_m(z=0,\Omega)=A^{\rm s}_m(z=0)\delta(\Omega)+B_m(z=0,\Omega).
    \label{Eq:SNsplit}
\end{equation}
Similarly, the total amplitude in mode $m$ at any point $z$ can be decomposed into the signal ($\Omega=0$) and noise ($\Omega\neq0$) amplitudes:
\begin{equation}
    A_m(z,\Omega)=A^{\rm s}_m(z)\delta(\Omega)+B_m(z,\Omega).
    \label{Eq:zSNsplit}
\end{equation}
As the light propagates down the fiber, both the signal and noise grow exponentially due to the linear optical gain $g$. Below the TMI threshold, signal amplitudes roughly grow independently of the noise and in the case of no gain saturation are given by:
\begin{equation}
    A^{\rm s}_m(z)=A^{\rm s}_m(0)e^{\frac{g}{2} z}e^{i\phi^{\rm NL}_m}.
    \label{Eq:Ssol}
\end{equation}
Here, $\phi^{\rm NL}_m$ is the nonlinear phase evolution in mode $m$. More importantly, noise also grows due to the transfer of power from the signal due to the thermo-optical scattering. The noise growth can be obtained by solving Eq.~\ref{Eq:CAS}, which is first simplified by retaining only the phase-matched terms, given by the condition $\beta_m-\beta_n+\beta_i-\beta_j=0$, leading to:
\begin{equation}
\begin{aligned}
\diff{ A_m(z,\Omega)}{z}  =&\frac{g}{2}A_m + ig \chi_0  \sum_{n} \int d\Omega' A_n(z,\Omega-\Omega')\\&  G_{mnnm}(\Omega') \int d\Omega^{''}  A_n^*(z,\Omega^{''})A_m(z,\Omega'+\Omega^{''}). 
\end{aligned}
\label{Eq:CASpm}
\end{equation}

\noindent
We have utilized $m=j, i=n$ as a solution to the phase-matching condition, reducing the triple sum in the second term of Eq.~\ref{Eq:CAS} to a single sum in Eq.~\ref{Eq:CASpm}. Note that $m=n, i=j$ is also a valid solution to the phase-matching condition, but such terms correspond to only a cross-phase modulation but no transfer of power as these terms result in only a uniform temperature shift ($q_{ij}\approx0$), thus we do not retain these terms. We have also assumed absence of any exact degeneracies. This assumption can break down especially in graded index fibers where significant degeneracies are present in which case the above equations needs to be modified slightly. For a detailed discussion see Supplementary Information Section I. Next, we substitute the ansatz for $A_m$ from Eq.~\ref{Eq:zSNsplit} and Eq.~\ref{Eq:Ssol} in Eq.~\ref{Eq:CASpm} and simplify the RHS to obtain the equation growth equation for $B_m(z,\Omega)$: 

\begin{equation}
    \frac{d B_m(z,\Omega)}{dz} = g(\frac{1}{2}+i\chi_0\sum_{n}G_{mnnm}(\Omega)P^{\rm s}_n(z))B_m(z,\Omega).
    \label{Eq:Nampgrowth}
\end{equation}
\noindent
All the terms which are quadratic or higher order in noise amplitudes are ignored. Each noise amplitude grows exponentially, independently of other noise amplitudes, and has contributions in growth rate from both linear amplification and nonlinear power transfer from the signal. $P^{\rm s}_m$ represents the signal power in mode $m$ and is given by ${|A^{\rm s}_m|}^2$ and $G_{mnnm}$ represents the contribution to noise growth in mode $m$ by signal power in mode $n$. Note that here we have considered a signal at a single frequency  without any linewidth broadening, since our goal is to study TMI in narrowband fiber amplifiers with a focus on coherent excitations. However, in some applications of fiber amplifiers signal linewidth broadening is utilized. In that case our theory can be straightforwardly generalized by directly solving Eq.~\ref{Eq:CASpm} instead of simplifying it to Eq.~\ref{Eq:Nampgrowth}. We can convert noise amplitude growth equations for mode $m$ into growth equations for noise power (given by $|B_m|^2$), by multiplying with $B_m^*$ on both sides and adding the complex conjugate term:
\begin{equation}
    \frac{d P^{\rm N}_m(z,\Omega)}{dz} = g(1+\sum_{n\neq m}\chi_{mn}(\Omega)P^{\rm s}_n(z))P^{\rm N}_m(z,\Omega).
    \label{Eq:NPgrowth}
\end{equation}
\noindent
Here, $P^{\rm N}_m$ denotes the noise power in mode $m$. The first term on the RHS in Eq.~\ref{Eq:Nampgrowth} corresponds to linear optical gain and each term in the second term corresponds to a transfer of power from a particular signal mode $n$ to noise in mode $m$ due to the thermo-optical scattering. We have ignored the self-coupling term\cite{Hansen2013theoretical} (corresponding to $G_{mmmm}$) since the grating formed by interference of mode $m$ with itself at a Stokes shifted frequency has a grating period ($2\pi/(q_{mm} \approx 0)$) much larger than the length of the fiber. For each mode pair $(m,n)$, we have defined a thermo-optical coupling coefficient $\chi_{mn}(\Omega)$, which is equal to $-2$ times the imaginary part of $G_{mnnm}$ multiplied with $\chi_0$. Eq.~\ref{Eq:NPgrowth} can be solved analytically, resulting in exponential growth in noise power in each mode $m$:
\begin{equation}
    P^{\rm N}_m(L,\Omega)=P^{\rm N}_m(0,\Omega)e^{gL}e^{G^{\tiny{\rm TMI}}_m(\Omega)}.
    \label{Eq:Nsol}
\end{equation}
$P^{\rm N}_m(0,\Omega)$ represents noise power in mode $m$ at the input end of the fiber. $P^{\rm N}_m(L,\Omega)$ is the noise power at the output upon exponential growth both due to the linear gain $g$ and the TMI gain which depends on signal power distribution: 

\begin{equation}
    G^{\tiny{\rm TMI}}_m(\Omega)=\sum_{n\neq m}\chi_{mn}(\Omega)\int_0^L gP_n^{\rm s}(z)dz
    \label{Eq:GTMI}
\end{equation}

\noindent
The $z$ integral can be simplified by using the signal growth formula given in Eq.~\ref{Eq:Ssol}, $\int_0^L gP_n^{\rm s}(z)dz =P_n^{\rm s}(L)-P_n^{\rm s}(0)\approx \Delta P \Tilde{P}^{\rm s}_n$, where $\Delta P$ is the total signal power extracted from the amplifier and  $\Tilde{P}^{\rm s}_n$ is the fraction of signal power in mode $n$ ($\sum_n \Tilde{P}_n^{\rm s}=1$), which is determined by the input excitation. Thus, TMI gain in mode $m$ is given by:

\begin{equation}
    G^{\tiny{\rm TMI}}_m(\Omega)=\Delta P \sum_{n\neq m}\chi_{mn}(\Omega)\Tilde{P}_n^{\rm s}\equiv \Delta P \bar{\chi}_m(\Omega).
    \label{Eq:GTMIsimp}
\end{equation}
\noindent
This formula suggests a relatively straightforward interpretation. The TMI gain for any mode depends on the total extracted signal power and effective thermo-optical coupling coefficient $\bar{\chi}_m(\Omega)$, which is given by the weighted sum of thermo-optical coupling coefficients with all the other modes, with weights depending on the input excitation. For a two-mode fiber with FM-only excitation ($\Tilde{P}^{\rm s}_1=1$), the above formula reduces to the well known formula for TMI gain\cite{dong2013stimulated,Hansen2013theoretical}, where it is given by the product of extracted signal power and the thermo-optical coupling between the two modes $\chi_{21}$. More generally, Eq.~\ref{Eq:GTMIsimp} can be used to derive the formula for the TMI threshold under arbitrary multimode excitations, as shown in the next section.

\subsection{Multimode TMI Threshold}

\begin{figure*}
    \centering
    \includegraphics[width=0.95\textwidth]{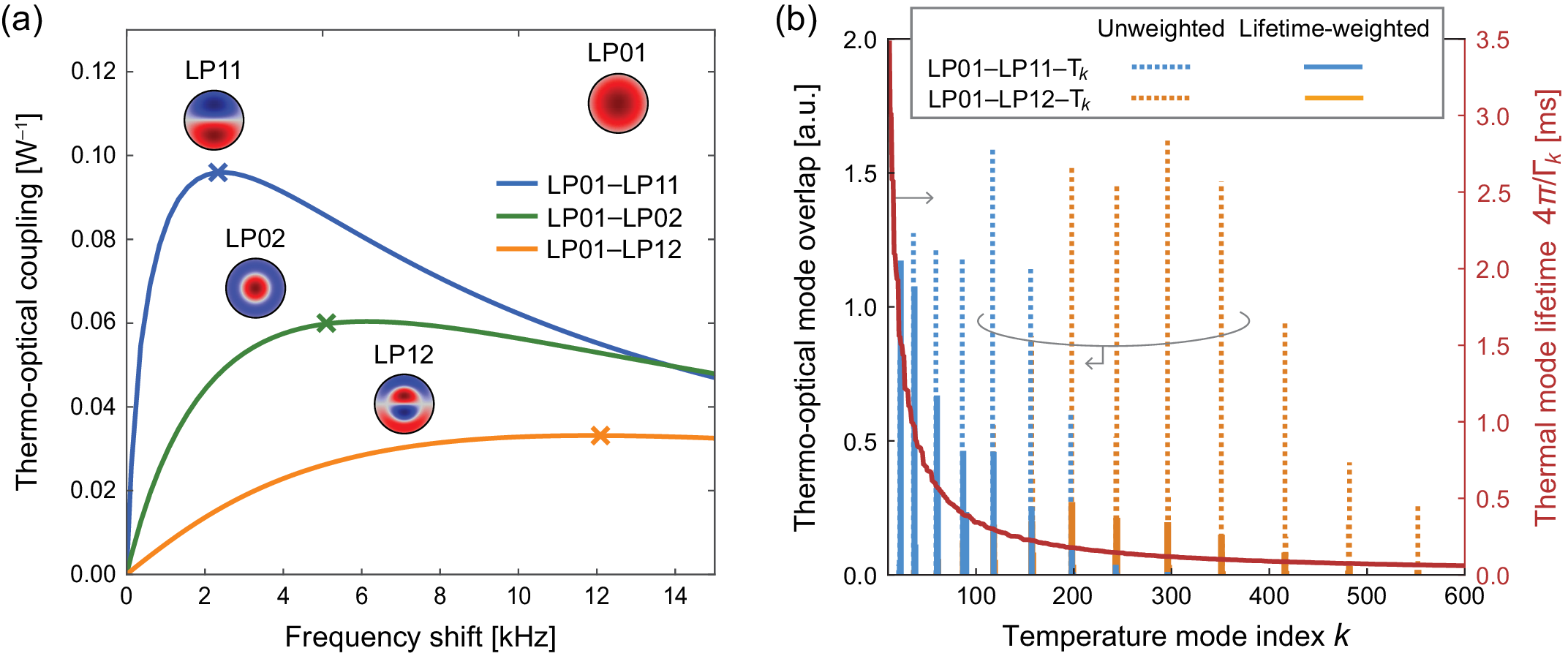}
    \caption{(a)Thermo-optical coupling coefficient between the FM (LP01) and three HOMs in increasing mode order (LP11, LP02, LP12) for circular fiber . Relevant mode profiles are shown in the insets. The thermo-optical coupling coefficient vanishes for zero Stokes shift ($\Omega=0$) and peaks at frequencies on the order of few KHz. LP01 couples most strongly with LP11 (blue curve) and the peak coupling decreases with increasing transverse spatial frequency mismatch between the modes (green and orange curves). The peak frequency and linewidth also increases with mode order of HOMs. (b) thermo-optical mode overlap for two different optical mode pairs: LP01-LP11 (blue) and LP01--LP12 (orange) with various temperature eigenmodes. The temperature modes are ordered by increasing eigenvalues and decreasing mode lifetimes, $2\pi\Gamma_k^{-1}$ (shown by red curve). For each optical mode pair only a few temperature eigenmodes have significant overlap (dotted bars), the ones which have spatial frequency similar to the frequency of interference pattern of optical modes within the fiber core. For neighboring optical mode pair (LP01--LP11), significant overlap occur with lower order temperature eigenmodes (with high mode lifetime) and for mode pair with relatively larger separation (LP01--LP12), significant overlap occurs with relatively higher order temperature eigenmodes (with short mode lifetimes). Note that the thermo-optical coupling coefficients are decreased when the mode lifetimes $4 \pi \Gamma_k^{-1}$ are smaller. Consequently, optical mode pairs with larger separation have weaker overall coupling as indicated when the thermo-optical overlaps are weighted by lifetime (solid bars), leading to the lower peak coupling and higher peak frequency as shown in (a).}
    \label{fig:Fig4}
\end{figure*}

The TMI threshold is defined as the total output power when the noise power becomes a significant fraction ($\xi>1\%$) of the signal power, which leads to the onset of a fluctuating beam profile\cite{Hansen2013theoretical,dong2013stimulated}. For a given amount of input noise power, this occurs when the highest TMI gain across all frequencies and modes becomes large enough. Therefore, a quantitative condition for the threshold can be derived from Eq.~\ref{Eq:Nsol} and Eq.~\ref{Eq:GTMIsimp}:

\begin{equation}
   P^{\rm N}(L)=P^{\rm N}(0)\:\:e^{gL}e^{(P_{\rm th} - P^{\rm s}(0))\bar{\chi}}=\xi P^{\rm s}(0) e^{gL}, 
    \label{Eq:TMIth0}
\end{equation}
which can be rearranged as:
\begin{equation}
    (P_{\rm th} - P^{\rm s}(0))\bar{\chi} = \log(\frac{\xi P^{\rm s}(0)}{P^{\rm N}(0)}), 
    \label{Eq:TMIth}
\end{equation}
\noindent
where, $P_{\rm th}$ is the TMI threshold, $L$ is the length of the fiber, $P^{\rm s}(0)$ and ${P^{\rm N}(0)}$ are the input signal and noise powers, respectively. We have introduced an overall thermo-optical coupling coefficient $\bar{\chi}$ which is equal to the maximum value of the effective thermo-optical coupling coefficients across all the modes and Stokes frequencies:

\begin{equation}
    \bar{\chi} = \max_{\Omega,m}\sum_{n\neq m}\chi_{mn}(\Omega)\Tilde{P}_n^{\rm s}.
    \label{Eq:X}
\end{equation}

\noindent
A key insight from the multimode TMI threshold formula in Eq.~\ref{Eq:TMIth} is that TMI threshold is roughly inversely proportional to the overall thermo-optical coupling coefficient $\bar{\chi}$, which depends on both the fiber properties through $\chi_{mn}$ and the input power distribution, ${\Tilde{P^{\rm s}_n}}$. It also depends weakly (logarithmically) on the input noise power and the noise fraction $\xi$ at which the threshold is set. Also, any increase in input signal power (or, seed power) leads to a corresponding increase in the TMI threshold. Typically the input power is significantly smaller than the TMI threshold, thus the overall relative change due to a seed power increase is small\cite{tao2018comprehensive,dong2022accurate}. Note that in Eq.~\ref{Eq:TMIth0} we have approximated the total Stokes power by the Stokes power in the mode with maximum Stokes gain at  Stokes frequency with peak gain. This is justified by the exponential nature of the Stokes growth. For a detailed derivation, see Supplementary Information Sec.II.  It can be easily verified that when only the fundamental mode is excited ($\Tilde{P}_1 = 1, \Tilde{P}_{n\neq1}=0$), the overall thermo-optical coupling coefficient, $\bar{\chi}$ simplifies to $\chi_{21}$, reducing our multimode TMI threshold formula to a previously derived formula in studies which only consider single mode excitation\cite{dong2013stimulated}. More generally, multimode excitation provides a parameter space to control the overall thermo-optical coupling coefficient and the TMI threshold, even for a fixed fiber. The TMI threshold is highest for the signal power distribution which minimizes the maximum effective thermo-optical coupling coefficient. The amount of tunability therefore depends on the properties of the pairwise thermo-optical coupling coefficients which we discuss in detail in the next section.

\subsection{Thermo-optical coupling}
\label{sec:TMIcoup}

The thermo-optical coupling coefficient between any modes $m$ and $n$ is proportional to the imaginary part of a  particular component of the Green's function, $G_{mnnm}$, and is equal to:

\begin{equation}
   \chi_{mn}(\Omega) = \chi_0\sum_{k}|\langle\vec{\psi}_m^*\cdot\vec{\psi}_n \Tilde{T}_k \rangle|^2\frac{D\Omega}{\Gamma_k^2+\Omega^2}.
   \label{Eq:chi}
\end{equation}

\begin{figure*}[t!]
    \centering
    \includegraphics[width=0.9\textwidth]{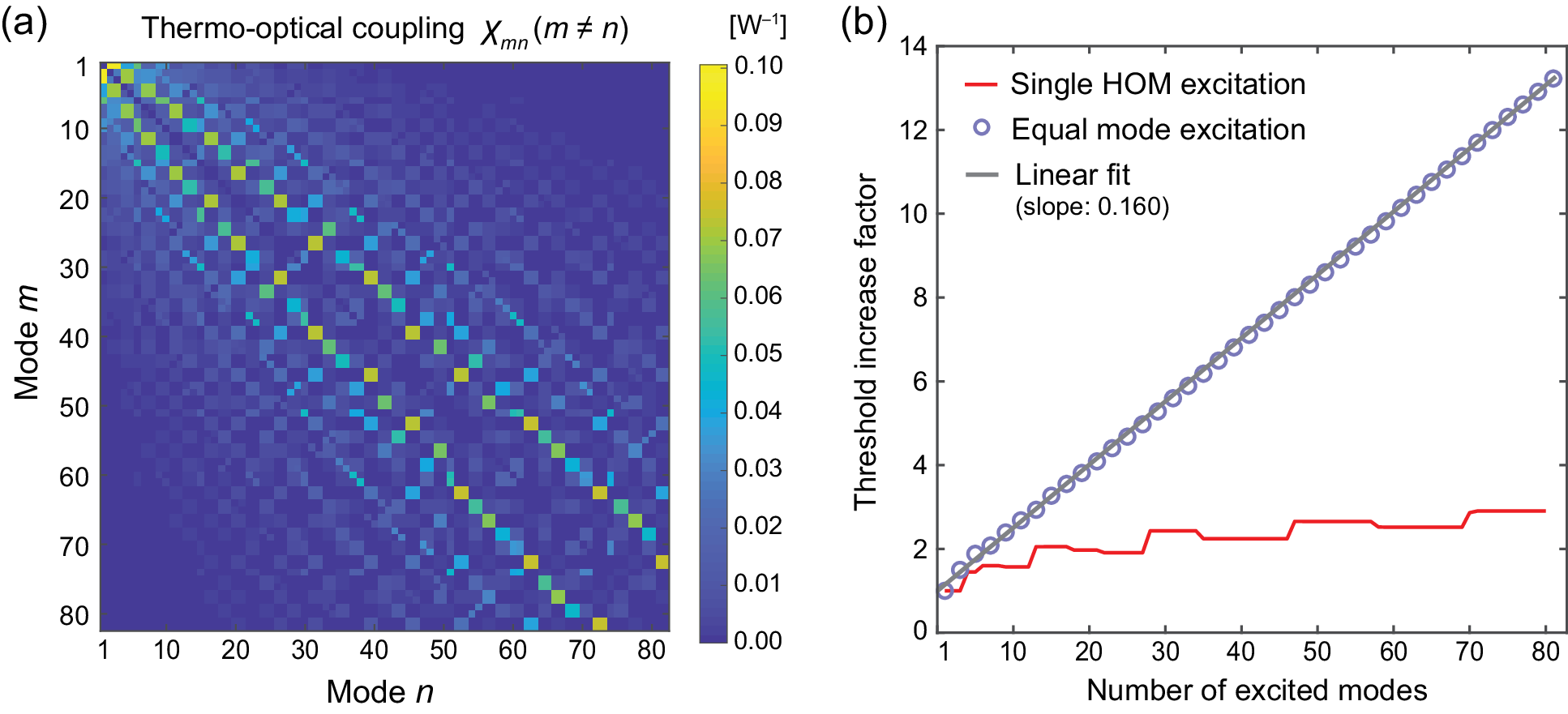}
    \caption{	(a) Thermo-optical coupling matrix for circular fiber  (Circular cross-section). Each element ${m,n}$ is a positive entry given by the peak value of thermo-optical coupling coefficient $\chi_{mn}(\omega)$. Self-coupling ($m=n$) is not considered for the reasons discussed in the phase-matching subsection (Sec.IIB). Only a few entries close to the diagonal have significant value, giving a banded matrix. (b) TMI threshold scaling in circular fiber  with the number of equally excited modes $M$. Threshold increase factor is defined by taking the ratio of TMI threshold with the threshold for FM-only excitation. As the number of excited modes are increased, the TMI threshold increases linearly and can reach upto $13$ times higher when all $82$ modes are excited. In comparison, the TMI threshold increase upon single HOM excitation (red curve) is significantly smaller.} 
    \label{fig:Fig5}
\end{figure*}

\noindent
$\chi_{mn}$ has contributions from each temperature eigenmode $k$. The strength of each contribution is proportional to the integrated overlap of the dot product of the optical mode profiles $\vec{\psi}_m$ and $\vec{\psi}_n$ with the temperature mode profile $\Tilde{T}^k$. Each contribution has a characteristic frequency curve which peaks at a frequency given by inverse mode lifetime $\Gamma_k$, and the peak value is proportional to the thermal mode lifetime $4\pi\Gamma_k^{-1}$. To illustrate the key properties of the thermo-optical coupling coefficient we calculate $\chi_{mn}(\Omega)$ for all the mode pairs of a circular step-index fiber with core diameter of 40 $\mu$m, cladding diameter of 200 $\mu$m and numerical aperture $NA=0.15$, supporting 82 modes per polarization. Detailed fiber or optical parameters used for calculation are provided in Table~\ref{tab:my_label}. We consider optical modes linearly polarized (LP) along the $x$-direction. Each LP mode is designated by two indices $[u,v]$ which correspond to the number of azimuthal and radial nodes
respectively\cite{okamoto2021fundamentals,qian1986lp} (See Fig.~\ref{fig:circ_modes}a). The modes are arranged in the order of decreasing longitudinal propagation constants. For instance, first five modes (excluding rotations) are LP01 (FM), LP11, LP21, LP02, and LP12. In our notation, therefore, $\chi_{12}\equiv \chi_{LP01,LP11}$. Note that LP modes are only approximate modes of the fiber, which become inaccurate for fibers with long lengths, in which case the exact vector fiber modes (HE and EH modes) need to be considered\cite{okamoto2021fundamentals}. For the purpose of our discussion we assume that LP modes are accurate enough for the length of the fiber considered. 
\begin{table}[b]
    \centering
    \begin{tabular}{|c|c|}
 \hline
 Parameter & Value \\ [0.5ex] 
 \hline\hline
 Core shape & Circular \\
 \hline
 Core diameter [$\mu$m] & 40 \\ 
 \hline
 Cladding diameter [$\mu$m] & 200\\
 \hline
 Core refractive index & 1.458\\
 \hline
Cladding refractive index & 1.45  \\
 \hline
Signal wavelength, $\lambda_{\rm s}$ [nm] & 1032 \\
\hline
Pump wavelength, $\lambda_{\rm p}$ [nm] & 976 \\
\hline
Number of optical modes & 82\\
\hline
Thermo-optic coefficient, $\eta$ $[{\rm K}^{-1}]$& $3.5\times 10^{-5}$\\
\hline
Thermal conductivity, $\kappa$ $[{\rm Wm}^{-1}{\rm K}^{-1}]$& $1.38$\\
\hline
Diffusion constant $D$ $[{\rm m}^{2}{\rm s}^{-1}]$ & $8.46\times 10^{-7}$\\
\hline
    \end{tabular}
    \caption{Detailed parameters for circular fiber.}
    \label{tab:my_label}
\end{table}

In Fig.~\ref{fig:Fig4}a we have shown the thermo-optical coupling coefficient between the FM (LP01) and three HOMs in increasing mode order (LP11, LP02, LP12) for this fiber. Relevant mode profiles are shown in the inset. The thermo-optical coupling coefficient in all three cases is zero at $\Omega=0$ and attains its peak value at frequencies on the order of few kHz,  consistent with the millisecond timescale of TMI\cite{Hansen2013theoretical,dong2013stimulated,Cesar2020transverse}. The FM couples most strongly with the lowest HOM (blue curve), and the peak coupling decreases with increasing transverse spatial frequency mismatch between the modes (green and orange curves). The peak frequency and linewidth are higher for larger spatial frequency mismatch between the modes. These results can be understood by investigating the specific temperature eigenmodes facilitating the coupling for different mode pairs, as shown in Fig.~\ref{fig:Fig4}b. The contribution of each temperature eigenmode to the coupling between a pair of optical modes is proportional to (1) the thermo-optical modal overlap integral and (2) the corresponding thermal mode lifetime $4 \pi\Gamma_k^{-1}$ (Eq.~\ref{Eq:chi}). Overall, only a few temperature eigenmodes, which match the transverse intensity profile of the interference between the optical modes, have a significant overlap integral.  For $\rm LP01-LP11$ coupling, the optical interference pattern has relatively lower spatial frequencies and thus has significant overlap with relatively lower order temperature eigenmodes (shown by blue dotted bars in Fig.~\ref{fig:Fig4}b). On the other hand, the LP01--LP12 mode pair has a significant overlap with relatively higher order temperature eigenmodes (shown by orange bars in Fig.~\ref{fig:Fig4}b). The thermal mode lifetime is shorter for temperature eigenmodes with higher transverse spatial frequency (as shown by red curve in Fig.~\ref{fig:Fig4}b), so the contribution of higher-order temperature eigenmodes is damped. Such intrinsic dampening of high spatial frequency contributions to the temperature are characteristic of the heat propagation being a diffusive process\cite{seyf2016method}. Consequently the lifetime weighted thermo-optical mode overlap for neighboring optical modes (blue solid bars) is much higher than for modes with larger separation of transverse spatial frequency (orange solid bars). As a result, peak thermo-optical coupling coefficient decreases with increasing spatial frequency mismatch between the optical modes.  
The quantitative relation between the spatial frequency mismatch and the lowered thermo-optical coupling coefficient is made possible our use of thermal eigenmodes (each with a definite spatial frequency) for expressing the temperature fluctuations.

This is a generic result for all optical mode-pairs. To explicitly show this, we calculate the peak value of $\chi_{mn}(\Omega)$ for all $\sim 10^4$ mode pairs in circular fiber . The resulting coupling matrix is shown in Fig.~\ref{fig:Fig5}a as a false color image. We have omitted the self coupling ($m=n$), since it is not responsible for intermodal power transfer between the modes\cite{Hansen2013theoretical}. Clearly, every mode couples strongly with only a few modes resulting in a highly banded/sparse coupling matrix. The strong coupling occurs when the spatial frequencies of two modes are closely matched, i.e., similar number of radial and azimuthal nodes ($\Delta u, \Delta v \leq 1 $). As we will see below, this banded nature of the coupling matrix is what allows a significant suppression of TMI upon multimode excitation.

\subsection{Threshold Scaling}
\label{sec:Thscaling}
Recall that the TMI threshold is defined as the output signal power at which the noise power in any mode becomes a significant fraction ($>1\%$) of the output signal power\cite{dong2013stimulated}, and the beam profile fluctuates dynamically rendering it useless for many applications\cite{Cesar2020transverse}. According to Eq.~\ref{Eq:TMIth} and Eq.~\ref{Eq:X}, the TMI threshold is inversely proportional to the overall thermo-optical coupling coefficient $\bar{\chi}$ which is a sum of $\chi_{mn}$ weighted by the fraction of signal power in each mode, which depends on the input excitation. Most previous TMI suppression efforts consider exciting only the fundamental mode and avoid sending power in HOMs, which our results indicate is not the optimal approach. To show this explicitly we consider two special cases of input excitation --- (1) FM-only excitation: all of the signal power is present in the FM ($\Tilde{P}_1^{\rm s}=1$) and noise is present in HOMs, and (2) Equal mode excitation : the input power is divided equally in $M$ modes of the fiber ($\Tilde{P}_n^{\rm s}=1/M$). We use Eq.~\ref{Eq:X} to calculate $\bar{\chi}$ in the two cases and use it to compare the TMI threshold. In the first case, the noise in the first HOM ($m=2$) has the highest growth rate due to signal power in the FM ($m=1$), giving $\bar{\chi}=\chi_{21}$. Since $\chi_{21}$ is the highest coupling out of all mode pairs (Fig.~\ref{fig:Fig5}a), FM-only excitation actually leads to the lowest TMI threshold. In the second case, all the modes contribute to the overall coupling to a given mode (say, $m=2$) but with weights reduced by a factor of $M$ i.e., $\bar{\chi}=\sum_{n\neq2}\chi_{2n}/M$. Due to the banded nature of $\chi_{mn}$ only a few elements in the sum are significant, leading to $\bar{\chi}\approx{s\chi_{21}}/{M}$. Here, $s$ is used to denote the average number of significant elements in the any row of the thermo-optical coupling matrix, which does not scale with $M$ and is roughly equal to 6 for circular fiber. Thus, for a highly multimode excitation ($M \gg s$), the TMI threshold can be significantly higher than the FM-only excitation. In fact, our reasoning predicts that the TMI threshold will increase linearly with the number of equally-excited modes, $M$. To verify this explicitly, we calculate the TMI threshold as $M$ is increased. We define the threshold increase factor (TIF) as the ratio of TMI threshold for $M$-mode excitation with that for FM-only ($M \approx 1$) excitation. Fig.~\ref{fig:Fig5}b shows the values of TIF as $M$ is varied. As predicted, the TMI threshold increases linearly with number of excited modes with a slope roughly equal to $1/{\rm s}$ $(\approx 0.16)$. When all 82 modes in the circular fiber are excited equally, we find a remarkable 13$\times$ enhancement of the TMI threshold over FM-only excitation. This confirms that highly multimode excitation in a multimode fiber amplifier can be a great approach for achieving robust TMI suppression. 

Our model and formalism is equally efficient for considering fibers with non-circular cross-sections, such as the D-fiber with more complex spatial structures of its modes, due to the underlying chaotic ray dynamics, as discussed and depicted in Fig. 3.  The physical argument for the suppression of TMI via highly multimode excitation does not rely on any special features of the fiber transverse modal geometry, and should be valid for general step-index fiber geometries. Confirmation of this expectation is shown in the Appendix A, where we find linear scaling of the TMI threshold with the number of modes excited for the D-fiber. The slope of the scaling line does depend weakly on both the geometry of the core and the cladding, but rather weakly, as discussed in the Appendix \cite{appendixA}

It should be noted that although we have only considered equal-mode excitation to achieve a higher TMI threshold, it is by no means a strict condition. Our theory predicts that multimode excitation will generically lead to a higher threshold than the FM-only excitation, due to the sparse nature of the thermo-optical coupling matrix.Therefore, it is expected to be a universal result, independent of the details of fiber composition, geometry (see above) and the precise distribution of excited modes. The level of enhancement will depend mainly on the effective number of excited modes. While most previous studies of TMI consider FM-only excitation, an increase in TMI threshold due to multimode excitation has been observed in some cases. A recent study by the current authors demonstrated an increase in TMI threshold upon multimode excitation using explicit time-domain numerical simulations of coupled optical and heat equations in 1-D cross-section waveguides\cite{chen2023suppressing}. Additionally, several recent experimental studies on the effect of fiber bending on TMI threshold in few-mode fibers provide evidence for our
predictions\cite{zhang2019bending,wen2022experimental,li2023mitigation}. They have observed that upon launching light in a fiber that supported multiple modes, a higher TMI threshold is obtained for a larger bend diameter (i.e., a loosely coiled fiber). The bend-loss of HOMs decreases when the bend diameter is larger, and thus the effective number of excited modes is higher, which leads to a higher TMI threshold in accordance with our predictions. It should be noted that these experimental findings are opposite to the predictions of previous theories which only consider FM excitation and predict that decreased HOM loss upon increasing bend diameter should lead to lowering of the TMI threshold\cite{dong2023transverse,tao2016suppressing}. The authors of these studies\cite{wen2022experimental,zhang2019bending} pointed out this anomaly and tried explaining these results by arguing that lower mode mixing leads to a higher threshold. In our opinion, this discrepancy is a result of ignoring the signal power in HOMs which is fully taken into account in our formalism and can straightforwardly account for their results.

\section{Nonlinear Model: Gain Saturation and Pump Depletion}

In the previous section, we approximated the signal growth by considering a simplified treatment of optical gain due to stimulated emission by assuming a constant gain coefficient $g_0$. This model was treated first to highlight the physics of multimode excitation and its impact on the thermo-optical coupling and TMI threshold. However, such a model of fiber gain neglects important effects such as gain saturation~\cite{siegman1986lasers,Smith2013increasing}, mode dependent gain/loss~\cite{ho2011mode} and pump depletion~\cite{dong2022accurate} which are present in any real high power fiber amplifier. In this section, we utilize a more realistic model of gain in multimode fiber amplifiers including all the above-mentioned effects and again are able to derive a more complicated, but still semi-analytical formula for thermo-optical coupling which gives the TMI threshold for arbitrary multimode excitations. This model should be used in order to make quantitative predictions to be compared with experiments.

We will show that the banded nature of the thermo-optical coupling matrix and the linear scaling of TMI threshold upon multimode excitation found for the simpler model in the previous section (\ref{sec:Thscaling}) are also found in the presence of gain saturation, mode dependent gain and pump depletion. This was expected, as the scaling of TMI threshold upon multimode excitation is a result of the diffusive nature of heat propagation leading to weak thermo-optical coupling between modes with large transverse spatial frequency mismatch. This mismatch of the spatial frequencies of the thermal and optical fluctuations is qualitatively unchanged by gain saturation and the other effects. We do find that including gain saturation impacts the exact value of TMI threshold, increasing it for all input excitations due to reduced dynamic heat load; this result is in agreement with the findings of previous studies of TMI in the case of single mode excitations which include these effects ~\cite{Smith2013increasing,hansen2014impact,Li2017experimental,dong2022accurate,li2023mitigation,naderi2013investigations,ward2016theory,tao2018comprehensive}. 

To set up our new model, we generalize the approach used by several previous studies to obtain TMI threshold upon single mode excitation~\cite{smith2013steady,hansen2014impact,Li2017experimental,dong2022accurate,li2023mitigation,naderi2013investigations,ward2016theory,tao2018comprehensive} for the case of multimode excitation. This involves two key steps. First, the signal amplitudes are obtained in each mode and along the entire fiber axis by numerically solving the saturated signal amplification equations coupled with the evolution of pump and the upper level population in the gain medium. In this step, any effect of Stokes wave (due to noise) and the thermo-optical coupling on the signal are neglected. This is consistent with the undepleted signal approximation utilized in the previous section, since as argued there, the backaction of the noise on the signal is negligible below the TMI threshold. In the second step, the signal amplitudes and upper level population are used to calculate relevant dynamic heat load and the resultant thermo-optical coupling is used to obtain the growth of Stokes power in each mode, which determines the TMI threshold.

\subsection{Signal Amplification}

We consider a co-pumped fiber amplifier with ytterbium (Yb) doped gain medium~\cite{paschotta1997ytterbium} in the fiber core, with a pump beam propagating in the pump core exciting the electrons to the upper level and creating inversion. As a result, the signal beam propagating in the fiber core undergoes amplification due to stimulated emission. The growth of signal amplitudes in each mode, $A^{\rm s}_m$, can be written as~\cite{smith2011mode}:

\begin{equation}
    \frac{dA^{\rm s}_m(z)}{dz} = \sum_n e^{i(\beta_n-\beta_m)z} g_{mn} (z) A^{\rm s}_n (z),
    \label{Eq:signalamp}
\end{equation}
where, $g_{mn}$ is the gain in mode $m$ due to amplitude in mode $n$ and is given by:
\begin{equation}
    g_{mn}(z) =  \frac{g_0}{2}\int d \vec{r}_\perp \frac{\psi^*_m (\vec{r}_\perp)\psi_n (\vec{r}_\perp)}{1+I_0 (\vec{r}_\perp,z)/I_{\rm sat} (z)}.
    \label{Eq:signalgain}
\end{equation}

\noindent
Here, the integral is over the doped area of the fiber cross-section. The new intensity-dependent factor in the denominator modifies the linear gain by taking into account the reduction in the inversion (gain saturation) due to removal of the pump energy by the signal amplification ~\cite{siegman1986lasers,paschotta1997ytterbium}; $g_0$ is the unsaturated gain coefficient. The amount of gain saturation depends on the ratio of the local signal intensity $I_0$ and a saturation intensity $I_{\rm sat}$. Both $g_0$ and $I_{\rm sat}$ depend on various properties of the gain medium and the local value of pump power, $P_p$ and their formulas are obtained from steady state rate equations~\cite{smith2011mode,cerjan2012steady} (for more details see Supplementary Information Section III). Notice that in addition to self-gain terms ($m=n$), we also must consider cross gain terms ($m\neq n$), which are now non-zero due to the spatially varying nature of the gain saturation term (referred to as spatial hole burning) and leads to non-linear mode coupling due to gain saturation ~\cite{siegman1986lasers,cerjan2012steady,smith2011mode}. Some previous studies of TMI, which consider irradiance based models~\cite{hansen2014impact,Li2017experimental}, ignore the spatial hole burning retaining only the self gain terms. For studying multimode excitations, including the hole burning is essential, since the multimode signal intensity will typically have rapid speckle-like spatial variations. Note that we have ignored any variation in the real part of the refractive index due to the gain saturation as this tends to be significantly smaller than the imaginary part for relevant wavelengths.

As the pump beam excites the gain medium it loses energy, leading to a decreasing pump power along the fiber axis, referred to as pump depletion~\cite{smith2011mode}. We take this into account quantitatively by introducing a standard equation for the evolution of pump power, which is given by~\cite{smith2011mode}:

\begin{equation}
\frac{dP_p(z)}{dz} = -g_p(z) P_p(z).
\label{Eq:pumpdecay}
\end{equation}

\noindent
The pump power decays exponentially with a longitudinally varying loss coefficient $g_p$ which depends on the local inversion and is given by~\cite{smith2011mode}:

\begin{equation}
    g_p(z) = \frac{N_{\rm Yb}}{A_{cl}} \int d\vec{r}_\perp (\sigma_p^a n_{\rm l}(\vec{r}_\perp,z) - \sigma_p^e n_{\rm u}(\vec{r}_\perp,z))
    \label{Eq:pumpgain}
\end{equation}

\begin{figure}[t!]
    \centering
    \includegraphics[width=0.45\textwidth]{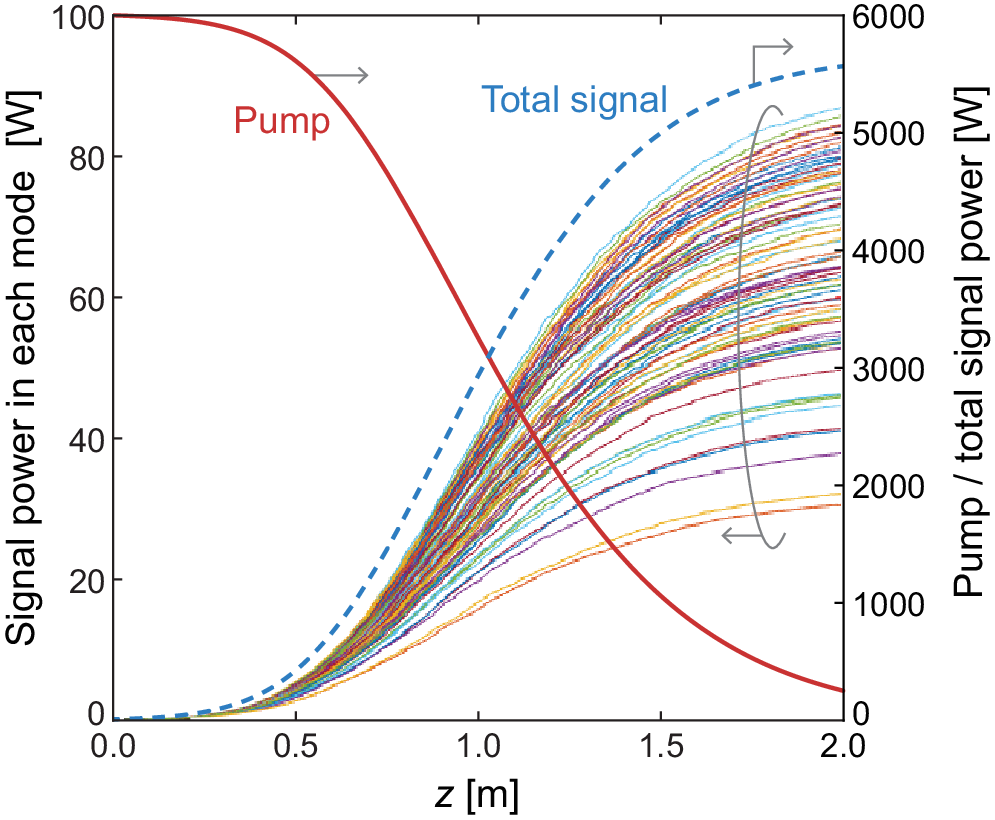}
    \caption{Pump power (solid red line, scale on the right y-axis) gets depleted as it is absorbed along the fiber creating inversion. This inversion leads to the growth of total signal power (shown by dotted red line, scale on the right y-axis) which grows exponentially at first and then linearly due to the gain saturation and eventually flattens out as most of the pump is depleted. All the other curves with various colors show the signal power in individual modes (scale on left y-axis) displaying the mode dependent gain due to spatial hole burning and differential overlap with the gain medium.}
    \label{fig:Fig6}
\end{figure}

\noindent
Here, the integral is over the doped area of the fiber cross-section. $\sigma_{\rm p}^{\rm e}$ and $\sigma_{\rm p}^{\rm a}$ are the emission and absorption cross-sections of the pump respectively, $N_{\rm Yb}$ is the density of doped ytterbium atoms and $A_{\rm cl}$ is the area of pump core which is typically same as the core plus the first cladding. Note that we are able to consider a single equation for pump power instead of a different one for each pump mode as in typical experiments the pump is incoherent and fills the fiber cross-section uniformly on average~\cite{paschotta1997ytterbium,Cesar2020transverse,zervas2017transverse}. The Yb doped gain medium we consider is a quasi-three-level medium. The pump photon causes a transition from the lower lasing level to the uppermost level after which the electron population immediately transitions downward to the upper lasing level causing heat generation in the process. This is known as quantum defect heating. $n_{\rm u}$ and $n_{\rm l}$ are the fraction of population in the upper and lower lasing levels respectively, satisfying $n_{\rm l}=1-n_{\rm u}$, where $n_{\rm u}$ is obtained by solving steady state rate equations, leading to~\cite{paschotta1997ytterbium,smith2011mode,naderi2013investigations,dong2022accurate,tao2018comprehensive}: 

\begin{equation}
    n_{\rm u} = \frac{\frac{P_{\rm p}}{A_{\rm cl}\hbar\omega_{\rm p}}\sigma_{\rm p}^{\rm a}+\frac{I_0}{\hbar\omega_{\rm s}}\sigma_{\rm s}^{\rm a}}{\frac{P_{\rm p}}{A_{\rm cl}\hbar\omega_{\rm p}}(\sigma_{\rm p}^{\rm e}+\sigma_{\rm p}^{\rm a})+\frac{I_0}{\hbar\omega_{\rm s}}(\sigma_{\rm s}^{\rm e}+\sigma_{\rm s}^{\rm a})+\frac{1}{\tau}}.
    \label{Eq:nu}
\end{equation}

Here, $\omega_{\rm s}$ and $\omega_{\rm p}$ are the signal and pump frequencies and $\sigma_{\rm s}^{\rm e}$ and $\sigma_{\rm s}^{\rm a}$ are the emission and absorption cross sections of the signal respectively. Each individual term involving scattering cross-sections is equal to either the rate of stimulated emission or absorption of the signal or the pump and has units of [s$^{-1}$]. $\tau$ is the upper state lifetime. $n_{\rm u}$ has a strong spatial dependence due to both the pump power $P_{\rm p}$ and the signal Intensity $I_0$.

A solution to the signal amplification equations (Eqs.~\ref{Eq:signalamp} and \ref{Eq:signalgain}) coupled with the evolution of the pump power (Eqs.~\ref{Eq:pumpdecay} and \ref{Eq:pumpgain}) and upper level population (Eq.~\ref{Eq:nu}) can be obtained by standard finite difference based numerical methods~\cite{leveque2007finite,laegaard2020multimode}. We utilize a centered difference approximation for all the $z$ derivatives and use it to iteratively update the values of the signal amplitudes and the pump power at $z+\delta z$ position from the values of respective gain coefficients at $z$ and values of the variables at $z-\delta z$ positions. At each $z$, the value of $g_p$, $g_{mn}$ and $n_{\rm u}$ are updated directly from Eqs.~\ref{Eq:signalgain}, \ref{Eq:pumpgain} and \ref{Eq:nu}. To validate our numerical model, we first simulated the 20/400 fiber amplifier studied in Smith et al.~\cite{smith2013steady} where single mode excitation was considered and obtained excellent agreement with their results. We also simulated the two-mode excitation case discussed in Li et al.~\cite{li2023mitigation} and found good agreement. 

Next we study the highly multimode step-index fiber amplifier discussed in section \ref{sec:TMIcoup}. All the parameters for Yb doped gain medium are taken from Table I in Smith et al.~\cite{smith2013steady}. The radius of the Yb doped area was considered to be equal to the fiber core radius. The unsaturated gain coefficient $g_0 \approx 4 \rm m^{-1}$, dopant concentration $N_{\rm Yb} = 3.25 \times 10^{25} \rm m^{-3}$ and length of the fiber L = 2 m. Initial pump power
depends on the TMI threshold for different excitations.For any given multimode input excitation we launch 10 W of seed power divided appropriately in the signal modes and a variable amount of pump power determined self consistently such that total output signal power is equal to the TMI threshold. The length of the fiber is chosen to be such that $>95\%$ pump power is converted to the signal power. Fig.~\ref{fig:Fig6} shows the results for the case when all 82 modes are excited equally at the input. 6 kW of pump power is launched in a 2 m long fiber and it decreases monotonically along the fiber axis as it excites the gain medium creating inversion. The signal power in each mode grows at first exponentially when the gain saturation is low and linearly when gain saturation becomes high and eventually flattening out when nearly all the pump is depleted. The final output power is roughly 5675 W leading to a  conversion efficiency of 0.945, which is close to the quantum efficiency limit ($\frac{\lambda_p}{\lambda_s}=0.9457$). Notice that the signal power in different modes grow with a different growth rate displaying mode dependent gain. This is a result of both the differential overlap with the fiber core where the gain medium is present and spatial hole burning induced variations in the fiber gain. At this stage thermo-optical effects have not yet been calculated.
In the next subsection, having obtained the saturated signal power along the entire fiber axis, we show how the gain-saturated thermo-optical coupling can be utilized to calculate the TMI threshold for arbitrary multimode excitations. 

\subsection{Thermo-optical coupling with gain saturation}
\begin{figure*}[t!]
    \centering
    \includegraphics[width=0.9\textwidth]{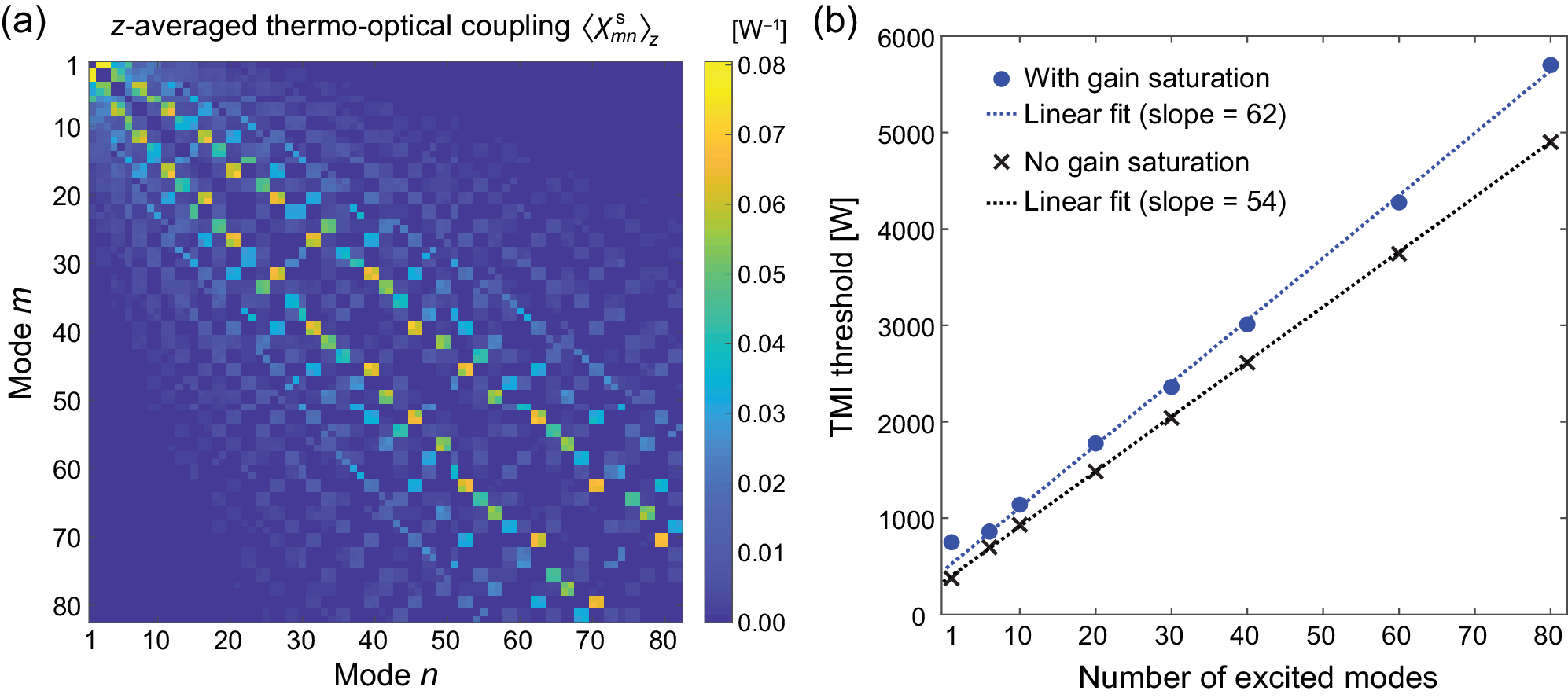}
    \caption{(a) z-integrated thermo-optical coupling matrix for fiber with circular cross-section including the effects of gain saturation and pump depletion. Self-coupling ($m=n$) is not considered for the reasons discussed in the phase-matching subsection (Sec.IIB). Similar to the unsaturated gain case, only a few entries close to the diagonal have significant value, giving a banded matrix. Saturation of the heat load leads to a reduction in overall thermo-optical coupling as displayed by the color-scale. (b) TMI threshold scaling in circular step-index fiber  with the number of equally excited modes $M$ with (blue circles) and without (black crosses) gain saturation. In both the cases the TMI threshold increases linearly with the number of modes. The dotted lines show best fit lines with slope = 54 W/mode without gain saturation and slope = 62 W/mode with gain saturation. When gain saturation is included the TMI threshold and the slope of linear scaling is higher. The difference in threshold in these two cases does not appear very dramatic because the y-axis spans a large range (0--6 kW) of values due to a large impact of multimode excitation. }
    \label{fig:Fig7}
\end{figure*}

The gain saturation impacts the local inversion and therefore also changes the amount of heat generated due to the quantum defect. The gain-saturated heat source is given by~\cite{hansen2014impact,dong2022accurate}:

\begin{equation}
    Q(\vec{r},t)= g_s q_{\rm D} I(\vec{r},t)=\frac{g_0 q_{\rm D} I(\vec{r},t)}{1+I(\vec{r},t)/I_{\rm sat}},
    \label{Eq:satheatSrc}
\end{equation}

\noindent
where, $g_{\rm s}$ and $g_0$ are the saturated and unsaturated gain coefficients and $q_{\rm D}$ is the quantum defect. The denominator comes directly from the saturation term in the gain coefficient and depends on the local signal intensity $I(\vec{r},t)$. At any point, the signal intensity can be written as a sum of a static contribution $I_0(\vec{r})$ and a dynamic contribution $\Tilde{I}(\vec{r},t)$ i.e. $I(\vec{r},t)=I_0(\vec{r})+\Tilde{I}(\vec{r},t)$. The first term is a result of self interference of the electric field at the signal frequency $\omega_0$, whereas the second term results from the interference between the signal at $\omega_0$ and the noise at Stokes shifted frequencies $\omega_0-\Omega$. Consequently, the heat profile has both a static and a dynamic contribution, $Q(\vec{r},t)=Q_0(\vec{r})+\Tilde{Q}(\vec{r},t)$. It is the latter term that results in the dynamic power transfer responsible for TMI . Below the TMI threshold the Stokes power is much smaller than the signal power, and hence the dynamic part of the intensity fluctuation is significantly smaller than the static part. Therefore, we can simplify the expression for $\Tilde{Q}$ by using the Taylor expansion in terms of $\Tilde{I}/(I_0+I_{\rm sat})$ and keeping the leading order term, giving the following expression~\cite{hansen2014impact} (for a detailed derivation, see Supplementary Information Section III):

\begin{equation}
    \Tilde{Q}(\vec{r},t) \approx  \frac{g_0 q_{\rm D} \Tilde{I}(\vec{r},t)}{{(1+I_0(\vec{r})/I_{\rm sat})}^2}.
    \label{Eq:satdynamicheatSrc}
\end{equation}

The dynamic heat source is proportional to the time varying intensity $\Tilde{I}$ and is inversely proportional to the square of the saturation term, which depends on local static intensity $I_0$. This quadratic behavior of the saturation term is a result of the extra contribution to the gain saturation from the dynamic part of the intensity, which comes with a negative sign upon the Taylor expansion, and partially cancels the direct term proportional to $\Tilde{I}$.\cite{hansen2014impact} Such a quadratic saturation of the dynamic heat load explains the well-known increase in the TMI threshold due to the gain saturation~\cite{smith2011mode,li2023mitigation} and was first pointed out by Hansen et al.~\cite{hansen2014impact} and later by several other authors\cite{dong2022accurate,Li2017experimental,zervas2017transverse}. 

Comparing the form of the relevant heat load in Eqs.~\ref{Eq:heatSrc} and \ref{Eq:satdynamicheatSrc}, we can directly derive a modified formula for the thermo-optical coupling $\chi^{\rm s}_{mn}$, taking into account the gain saturation:

\begin{equation}
\begin{aligned}
   \chi_{mn}^{\rm s}(\Omega,z) = \chi_0\sum_{k}\frac{D\Omega}{\Gamma_k^2+\Omega^2}&\int d\vec{r}_\perp\vec{\psi}_m^*\cdot\vec{\psi}_n \Tilde{T}_k \\ & \times \int d\vec{r}_\perp\frac{\vec{\psi}_m\cdot\vec{\psi}_n^* \Tilde{T}_k^*}{{(1+I_0(\vec{r}_\perp,z)/I_{\rm sat})}^2}.
   \label{Eq:chisat}
\end{aligned}
\end{equation}

\noindent
The key modification in the formula for thermo-optical coupling upon gain saturation is the presence of a saturation term in the overlap integral associated with the heat equation. Here, we have written the transverse integrals explicitly instead of using $\langle.\rangle$ as we did in Eq.~\ref{Eq:chi} to highlight the space-dependent nature of the saturation term due to the spatial-hole burning. In addition, since the local intensity $I_0$ in the denominator depends on $z$, now the thermo-optical coupling varies along the fiber axis. This modifies the form of the TMI gain given in Eq.~\ref{Eq:GTMI} by shifting $\chi^{\rm s}_{mn}$ inside the $z$ integral:

\begin{equation}
    G^{\tiny{\rm TMI}}_m(\Omega)=\sum_{n\neq m}\int_0^L\chi^{\rm s}_{mn}(\Omega,z) g_0 P_n^{\rm s}(z)dz.
    \label{Eq:GTMIsaat}
\end{equation}

\noindent
Here, $P^{\rm s}_n$ is the signal power in mode $n$ calculated with gain saturation in the first step discussed in the previous subsection. Note that in the above formula, the unsaturated gain coefficient $g_0$ is used, as the effect of saturation on the heat load is already considered in the thermo-optical coupling $\chi_{mn}^{\rm s}$. For a direct comparison with the form of the TMI gain given in Eq.~\ref{Eq:GTMIsimp}, we define a $z$-integrated thermo-optical coupling as follows:

\begin{equation}
    \langle \chi^{\rm s}_{mn}(\Omega)\rangle_z = \frac{g_0\int_0^L dz  \chi^{\rm s}_{mn} (z,\Omega) P^{\rm s}_n(z)}{P^{\rm s}_n(L) -P^{\rm s}_n(0)}. 
    \label{Eq:chizavg}
\end{equation}

Here the denominator is for normalization, chosen such that it is equal to $\int_0^L {\rm d}z g_{\rm s} P^{\rm s}_n(z)$, ensuring that in the absence of gain saturation the $z$-integrated $\chi$ becomes equal to the value of $\chi$ everywhere. This allows us to write an expression for TMI gain similar to the one in Eq.~\ref{Eq:GTMIsimp}:

\begin{equation}
    G^{\tiny{\rm TMI}}_m(\Omega)=\Delta P \sum_{n\neq m}\langle\chi_{mn}(\Omega)\rangle_z\Tilde{P}_n^{\rm s}.
    \label{Eq:GTMIsimpsat}
\end{equation}

The TMI gain in any mode is given by a product of the total extracted power $\Delta P$ and the weighted sum of $z$-integrated thermo-optical coupling with all the other modes. The weight, $\Tilde{P}^{\rm s}_n$ is the fraction of signal power extracted by mode $n$ ($\Tilde{P}^{\rm s}_n=\frac{P^{\rm s}_n(L) -P^{\rm s}_n(0)}{\Delta P}$) and can be controlled by the input excitation. 

Using the formulas in Eqs.~\ref{Eq:chisat} and \ref{Eq:chizavg} we were able to calculate the thermo-optical coupling for all the mode pairs in the 82-mode circular step-index fiber amplifier discussed in the previous section everywhere along the fiber when all 82-modes were equally excited and calculated the $z$-integrated coupling matrix in the presence of gain saturation. The results are shown in Fig.\ref{fig:Fig7}a. It can be seen that the thermo-optical coupling matrix remains sparse and approximately banded, even with gain saturation. This is as we expected from the physical arguments above. Although the structure of the matrix is unchanged, the exact values of the couplings are certainly impacted by the gain saturation. It can be seen by comparing the scale in Figs.~\ref{fig:Fig7}a and \ref{fig:Fig4}a that the $z$-integrated thermo-optical coupling is lowered due to the gain saturation for all the mode pairs by roughly $20 \%$. This leads to a higher value of TMI threshold upon including the gain saturation as we will see in the next subsection. The amount of coupling reduction depends on the ratio of the size of the pump core and the gain core as that determines the degree of gain saturation. For a larger pump core, the pump intensity in the gain core is lower decreasing the saturation intensity and increasing the amount of gain saturation, leading to a smaller value of z-integrated thermo-optical coupling. Note that the reduction in overall thermo-optical coupling also depends on the number of modes excited albeit weakly. 

\subsection{TMI Threshold scaling}

As just noted, gain saturation affects the local heat load and reduces the resultant thermo-optical coupling, leading to an increased value of TMI threshold, a fact that has been demonstrated both theoretically and experimentally for single mode excitations~\cite{smith2013steady,hansen2014impact,Li2017experimental,dong2022accurate,li2023mitigation,naderi2013investigations,ward2016theory,tao2018comprehensive}. To calculate the TMI threshold for multimode excitations we utilize the formula for TMI gain in Eq.~\ref{Eq:GTMIsimpsat}, which produces a similar formula for TMI threshold as given in Eq.~\ref{Eq:TMIth}. The only modification is that the overall thermo-optical coupling coefficient $\bar{\chi}$ is now given by a weighted sum of $z$-integrated thermo-optical coupling between various mode pairs:

\begin{equation}
    \bar{\chi} = \max_{\Omega,m}\sum_{n\neq m}\langle\chi^{\rm s}_{mn}(\Omega,z)\rangle_z\Tilde{P}_n^{\rm s}.
    \label{Eq:Xsat}
\end{equation}

The TMI threshold is inversely proportional to the overall thermo-optical coupling coefficient $\bar{\chi}$, which depends strongly on the input excitation via  signal power distribution in various modes $\{\Tilde{P}^{\rm s}_n\}$. As in the simpler model, due to the sparse nature of the saturated thermo-optical coupling matrix we expect power division to increase the TMI threshold. To show this quantitatively, we first calculate the TMI threshold when only the fundamental mode (FM) is excited and subsequently when various number of modes are equally excited. In each instance, the amount of pump power is chosen such that the total output signal power is equal to the TMI threshold. When  only the FM is excited, we obtain a TMI threshold equal to 775 W. To validate this result, we compare it to the predicted value of the TMI threshold in Smith et al.~\cite{Smith2013increasing} for a fiber with the same ratio of the diameter of the gain core and pump core (50/250) as in our fiber (40/200). A value of 785 W is obtained for the TMI threshold  for such a fiber from Table II in Ref.~\cite{Smith2013increasing}, which matches closely with our prediction. Without the gain saturation, the TMI threshold is found to be 375 W, which is also in close agreement with prediction in Ref.~\cite{Smith2013increasing} (345 W). In the previous studies on narrowband amplifiers involving large mode area (LMA) fibers (both step-index and photonic crystal fibers), the TMI threshold ranged from 200 W to 800 W \cite{Robin2014modal,kholaif2023impact,palma2023mitigation}. The TMI threshold upon FM-only excitation in this fiber is therefore comparable to LMA fibers.  As the number of modes excited are increased, the TMI threshold also increases linearly especially for large number of modes and reaches close to 5700 W when all 82 modes are equally excited (Fig.~\ref{fig:Fig7}b). The origin of this linear scaling is the same as in the simpler model.  The slope of the linear increase in TMI threshold is higher with saturation (62 W/mode) than without gain saturation (54 W/mode) since the starting point of the curve (fundamental mode threshold) is higher when gain saturation is included. The value of TMI threshold with gain saturation is higher by an amount ranging from 100 W to 800 W for various number of excited modes. For a fiber with a larger pump core or a smaller gain core this difference is expected to be even higher. As mentioned above, the reduction in thermo-optical coupling due to gain saturation which leads to increased threshold weakly depends on the number of modes excited. When gain saturation is taken into account, this effect can compete with linear enhancement in the TMI threshold for low number of modes causing deviations from strict linear scaling at the lower end in Fig.~\ref{fig:Fig7}b. For a large number of excited modes, linear enhancement dominates giving asymptotically linear scaling of TMI threshold with the number of modes excited.

\noindent

\section{Discussion and Conclusion}

In this work we have developed a theory of TMI, which can be used for efficient calculation of TMI threshold in narrowband fiber amplifiers for arbitrary multimode excitations and fiber geometries. A key result obtained from this theory is that TMI threshold increases linearly with effective number of excited modes. The scaling is a result of the diffusive nature of the heat propagation underlying the thermo-optical mode coupling. As such, it is expected to be applicable to a broad range of fibers. This opens up the use of highly multimode fibers as a promising avenue for instability-free power scaling in high-power fiber amplifiers.

The existing high-power multimode fiber amplifiers do not produce spatially coherent output that can be focused to a diffraction-limited spot or collimated to a Gaussian beam. By contrast, our approach allows the generation of spatially coherent beam out of a highly multimode fiber amplifier. This capacity will expand the potential applications of multimode fiber amplifiers, e.g., for a large-scale laser interferometer and coherent beam combining. It is sometimes mistakenly assumed that the presence of multiple spatial modes necessarily leads to a poor output beam quality and a high value of $M^2$.~\cite{zhan2009degradation,chu2019experimental} However, this belief is not correct, as previous work on the manipulation of coherent optical fields has shown~\cite{Florentin2017shaping,cizmar2011shaping,geng2021high}. Indeed the fields in multimode fibers can maintain a high output beam quality as long as the light remains sufficiently coherent~\cite{siegman1993defining,yoda2006beam}, i.e., the signal linewidth is narrower than the spectral correlation width of the fiber. In a typical scenario involving a 10-meter-long silica fiber with an NA of 0.1, the spectral correlation width is approximately 2 GHz for a laser wavelength of 1032 nm. In fact, by using an SLM to wavefront-shape the input light to a fiber, it is possible to obtain a diffraction-limited spot after coherent multimode propagation in both passive~\cite{gomes2022near} and active fibers~\cite{Florentin2017shaping}, which can be easily collimated using a lens. It is noteworthy that the SLM may introduce a few percent of optical loss due to diffraction by pixel edges into higher orders and absorption by the various layers in the SLM. However, since the SLM modulates only the input seed, its loss is negligible, as long as the amplifier operates in the gain-saturation regime where the output power depends mostly on the pump power.
Since focusing to diffraction-limited spot in the near field necessarily leads to excitation of multiple fiber modes, our theory predicts that it will lead to a higher TMI threshold than FM-only excitation, with a scaling proportional to the effective number of fiber modes. Hence the method just discussed can be used to increase the TMI threshold while maintaining good beam quality. A similar result was recently demonstrated experimentally for the case of Stimulated Brillouin Scattering (SBS) in multimode fibers; the threshold for the onset of SBS was demonstrated to increase significantly by focusing the output light of a multimode fiber to a diffraction-limited spot\cite{chen2023mitigating}.

Above we derived computationally tractable semi-analytic formulas for the TMI threshold in the presence of multimode excitation. First with a model that neglects gain saturation and related effects but illustrates the fundamental mechanism of suppression of TMI through the banded nature of the thermo-optical coupling matrix, leading to a linear threshold increase with the number of excited modes in the signal. Within this model all calculations are effectively linearized. Second, we improved the model, by calculating the saturated signal and pump depletion with a non-linear computational method, which is then used in a generalization of the first model to calculate the thermo-optical coupling matrix and the TMI threshold.  The model again finds a sparse coupling matrix and a linear scaling of the threshold with the number of excited modes. The absolute threshold is increased compared to the simpler model, as expected, but the difference is not dramatic.  This model incorporates all of the major effects missing in the first model but present in realistic fiber experiments: gain saturation and hole-burning of the signal, mode-dependent gain/loss and depletion of the pump; hence it is appropriate for quantitative comparisons with experimental data. 

We have still ignored any random linear mode coupling\cite{ho2011mode} in the fiber. This assumption can be relaxed relatively straightforwardly; the multimode TMI threshold is typically increased by the presence of mode mixing, since it promotes equipartition of signal power in various modes\cite{chen2023suppressing}. Note that our results are valid only under the assumption that the random linear mode coupling does not have a strong temporal variation on the time scales faster than what can be compensated by the use of spatial light modulators. This is typically true for most high quality multimode fibers.

Experimental validation of the theoretical results provided in this paper would be an important next step. Time-domain numerical simulations of optical and heat equations for up to 5 mode excitations in 1-D cross-section waveguides are provided in Ref.~\cite{chen2023suppressing} and are in good agreement with our theory. Additionally, several recent experimental studies investigating the effect of fiber bending on the TMI threshold in few-mode fibers provide evidence for our predictions\cite{zhang2019bending,wen2022experimental,li2023mitigation}. It has been observed that in a few-mode fiber, when the bend diameter of the fiber is increased (fiber being loosely coiled), a higher TMI threshold is obtained. For a large bend diameter, the bending loss of the HOMs decreases, thus the effective number of excited modes is higher, which leads to a higher TMI threshold in accordance with our predictions. In fact, the previous theories which only consider FM-only excitation predict that decreasing the HOM loss by increasing bend diameter should lead to lowering of the TMI threshold, in contrast to the experimental findings\cite{dong2023transverse,tao2016suppressing}. This discrepancy is a result of ignoring the signal power in HOMs, which is fully taken into account in our theory.  More systematic experimental studies are needed to investigate our predictions quantitatively.

In recent years, there has been a significant progress in fabricating low-loss fibers with non-circular cross-sections\cite{velsink2021comparison,ying2017recent}. These fibers have been proposed as a way to manipulate the strength of nonlinearities\cite{morris2012influence}. As noted above, the theory derived here can be used to calculate the TMI threshold for any fiber cross-section geometry. In Appendix A we have utilized it to demonstrate that a D-shaped fiber performs better than a standard circular fiber in raising the TMI threshold via multimode excitation. This interplay between input excitation and fiber geometry can provide a number of avenues for manipulating the strength of nonlinearities in the fiber. We believe that our theory can be utilized as a framework for customizing the strength of the thermo-optical nonlinearity using optimized fiber designs \cite{he2020machine}. Utilizing the spatial degrees of freedom of the fields to control nonlinear effects is becoming a standard tool, and has been demonstrated for diverse effects such as SBS\cite{chen2023mitigating,wisal2023theory,wisal2022generalized}, stimulated Raman scattering (SRS)\cite{tzang2018adaptive}, and Kerr nonlinearity\cite{wright2015controllable, krupa2017spatial}. Our work contributes to this exciting area by bringing the thermo-optical nonlinearity, with its different physical origin, into this category and at the same time providing a solution to the practical challenge of TMI suppression. 

\section*{Supplementary Material}

A supplementary material is attached for additional information. In the first section, we provide a detailed discussion on the phase matching for nonlinear thermo-optical scattering in multimode fibers. In the second section, we provide more details regarding the derivation of multimode TMI threshold formula and justify the use of only the Stokes mode with maximum gain at peak frequency. In section III, we provide a derivation of the saturated gain coefficient along with the formulae for various gain saturation parameters. In addition we derive the formula for saturated dynamic heat load. Finally, a quantitative comparison of our model is provided with previous studies on TMI in single mode and few mode fiber amplifiers. 

\begin{acknowledgments}
We thank Ori Henderson-Sapir, Heike Ebendorff-Heidepriem, and David Ottaway at The University of Adelaide, Stephen Warren-Smith and Linh Viet Nguyen at University of South Australia, and Peyman Ahmadi at Coherent for stimulating discussions. We acknowledge the computational resources provided by the Yale High Performance Computing Cluster (Yale HPC). This work is supported by the Air Force Office of Scientific Research (AFOSR) under Grant FA9550-20-1-0129. We also acknowledge the support of Simons Collaboration on Extreme Wave Phenomena Based on Symmetries. 
\end{acknowledgments}

\section*{author declaration}
The authors declare no conflict of interest.

\begin{appendices}
\appendix
\section{Impact of Fiber Geometry} \label{appendixA}

\begin{figure}[b]
    \centering
    \includegraphics[width=0.45\textwidth]{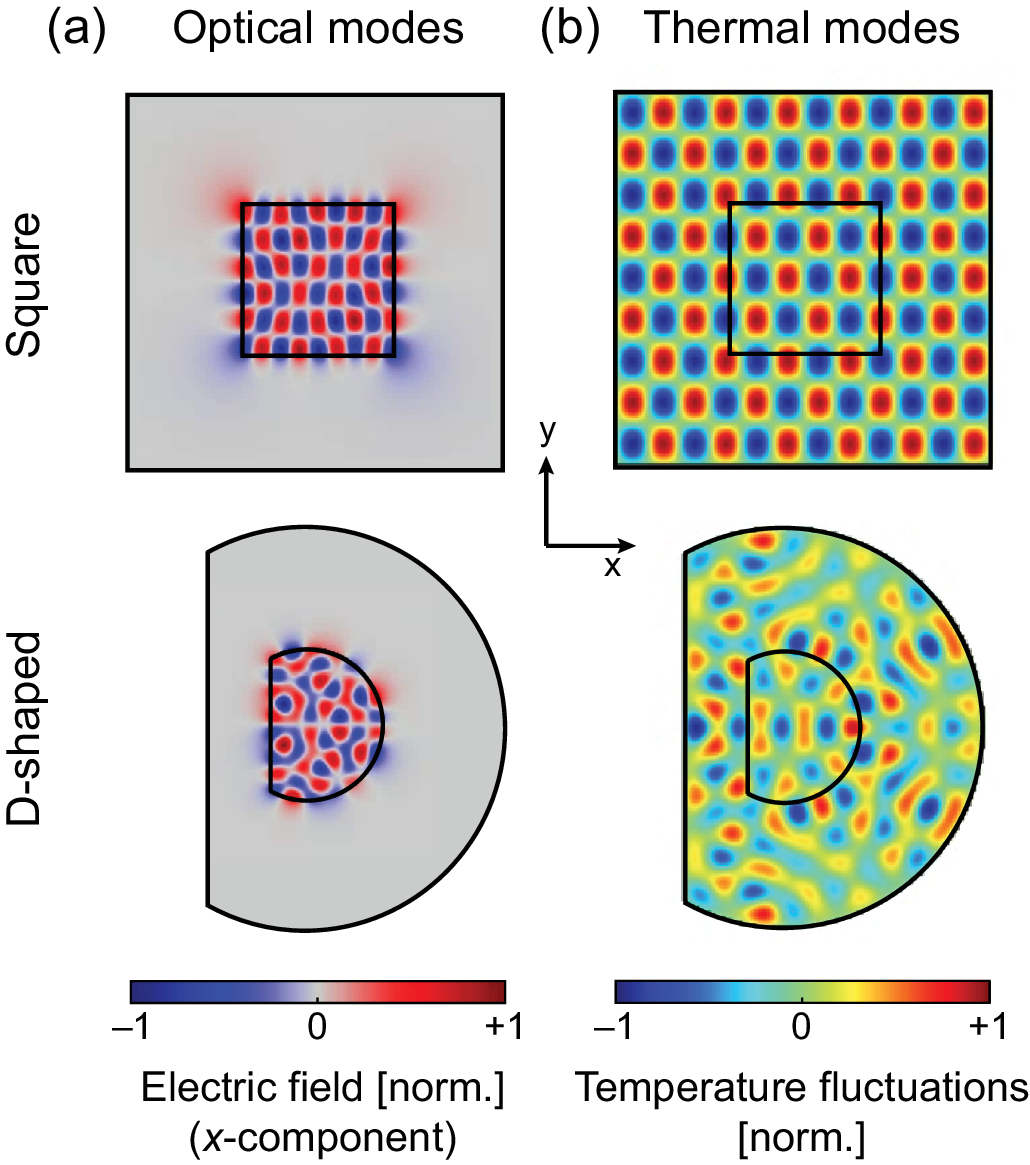}
    \caption{(a) Amplitude profiles of guided optical modes for D-shaped and square cross-section step-index fibers. The optical modes are guided in the core of the fiber. (b) Amplitude profiles for thermal modes for fibers with D-shaped and square shaped cladding. The thermal modes are the spatial eigenmodes of the heat equation with constant temperature at the cladding boundary. Each thermal mode fills the entire fiber cross section. Both optical and thermal modes in the square fiber are structured and have a particular number of nodes along x and y axes. The modes in  the D-shaped fiber have a random structure as D-shaped cavities are wave-chaotic.} 
    \label{fig:FigA1}
\end{figure}
\begin{figure*}[t!]
    \centering
    \includegraphics[width=0.99\textwidth]{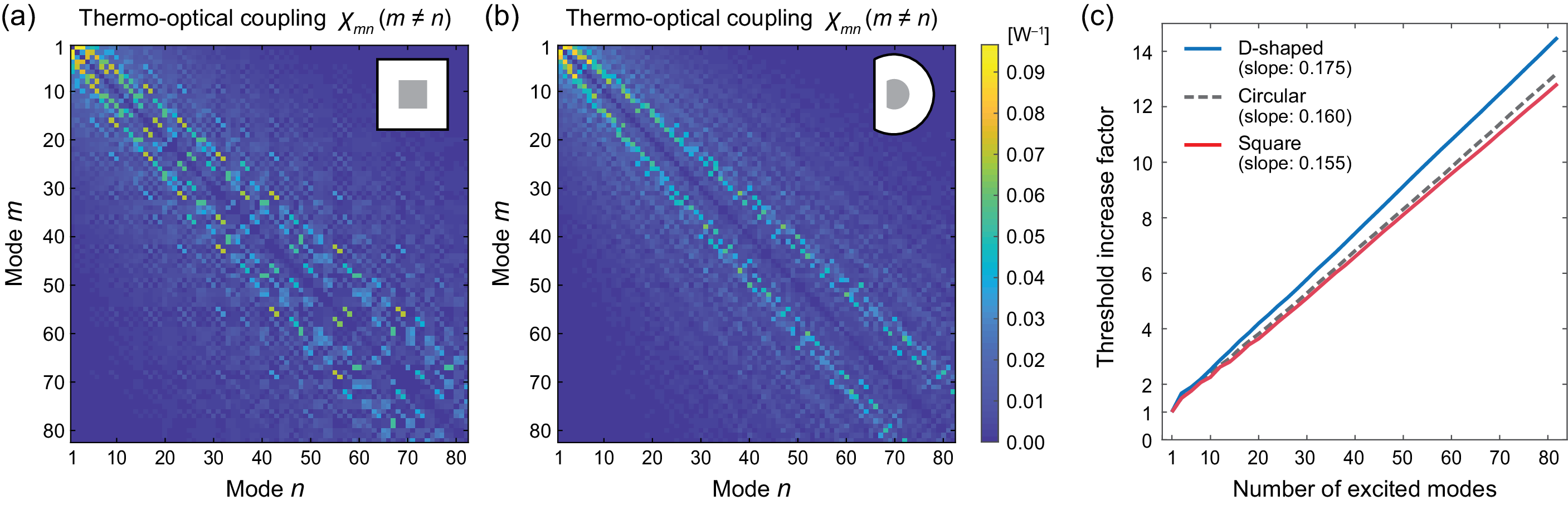}
    \caption{Thermo-optical coupling matrix for (a) square fiber (b) D-shaped fiber. Self-coupling is not considered ($m=n$). Similar to circular fiber, the thermo-optical coupling matrices for both square fiber and $C$ are also banded. (b) As a result TMI threshold in square fiber (red) and D-shaped fiber (blue) also increases linearly with the number of equally excited modes $M$. The slope of the scaling is highest for D-shaped fiber and lowest for square fiber. The scaling for circular fiber is reproduced as the dotted curve for reference.}
    \label{fig:FigA2}
\end{figure*}

In the main text, we considered a standard step-index multimode fiber with a circular cross-section, which shows that TMI threshold increases linearly with number of excited modes owing to the banded nature of the thermo-optical coupling matrix. This banded nature is a fundamental consequence of the diffusive nature of the heat propagation which underlies the thermo-optical scattering due to the intrinsic damping of high-spatial-frequency features. As a result, we expect the banded nature of the thermo-optical coupling matrix to be maintained even in fibers with non-standard geometries\cite{velsink2021comparison,ying2017recent}; while the number of significant elements in the coupling matrix, and their relative values can depend on the particular fiber geometry. As such, we expect the linear scaling of the TMI threshold to be maintained, but the slope to differ for fibers with different cross-sectional geometries.  Thus studying different fiber geometries to determine which yields a higher slope of threshold increase with $M$ may be useful in designing fibers with enhanced TMI thresholds.

As our formalism does not assume any particular fiber cross-section geometry, unlike most previous approaches\cite{dong2013stimulated,Hansen2013theoretical,zervas2017transverse}, it can be utilized straightforwardly for calculating the TMI threshold for different fiber geometries under multimode excitation. We consider two additional fiber cross-sectional geometries other than circular: square fiber and D-shaped fiber (referring to a circular shape truncated by removing a section bounded by a chord). For simplicity we ignore gain saturation and pump depletion in this section as these effects do not change the TMI threshold scaling qualitatively as shown in the main text. A reason to study these particular shapes is because square and D-shaped cross-sections support modes with statistically different profiles. While a square shape results in an integrable transverse "cavity", supporting modes with regular spatial structure\cite{marcatili1969dielectric}(as does the circular fiber); in contrast, the D-shaped cross-section leads to wave-chaotic behavior in the transverse dimensions, with highly irregular and ergodic modes\cite{bittner2020spatial}. Due to this property, D-shaped micro-cavities have found applications in suppressing instabilities and speckle-free imaging in 2D semiconductor micro-lasers\cite{bittner2018suppressing,redding2015low}. Similar to the circular fiber, the square fiber is chosen to have a core width of 40 $\mu$m and a cladding width of 200 $\mu$m. For the square fiber, we consider a slightly smaller NA of 0.135 as it has slightly larger core area compared to the circular fiber, such that it also supports 82 modes per polarization. In the D-shaped fiber, the core shape is formed by slicing a circle of diameter 40 $\mu$m with a line at a distance 10 $\mu$m (half the circle radius) from the centre. Similarly the cladding is obtained by slicing a circle of diameter 200 $\mu$m at the same relative distance. For the D-shaped fiber, we consider a slightly larger NA (0.17) as it has slightly smaller core area compared to circular fiber, such that it also supports 82 modes per polarization. All other relevant parameters for calculation are kept the same for all three fibers and are given in Table~\ref{tab:my_label}.

Similar to the circular fiber, we first calculate the thermo-optical coupling coefficients for all mode pairs for both the square fiber and D-shaped fiber by using the formula in Eq.~\ref{Eq:chi}. We calculate the optical modes in each case by using the Wave-Optics module in COMSOL\cite{comsol}. Example of an optical mode for each fiber is shown in Fig.~\ref{fig:FigA2}a. As expected, the optical modes for the square fiber are highly structured whereas optical modes for D-shaped fiber are irregular. We also calculate first $1000$ temperature eigenmodes for each fiber by using the Coefficient-Form-PDE module in COMSOL\cite{comsol}. The temperature eigenmodes depend on the cladding shape, which in this case is chosen to be same as the respective core shape. Example of a temperature mode for each fiber is shown in Fig.~\ref{fig:FigA1}b. The temperature eigenmodes have similar spatial properties to optical modes except they are spread out over both the core and cladding. We numerically evaluate the overlap integrals of the optical and temperature modes to find $\chi_{mn}(\Omega)$. We take the peak values over frequency for each mode pair to obtain effective coupling matrices, which are shown in Fig.~\ref{fig:FigA2}a and Fig.~\ref{fig:FigA2}b. As expected both fibers produce banded coupling matrices with a small number of significant elements. In each case $\chi_{21}$ is the largest element, suggesting FM-only excitation have the lowest TMI threshold and multimode excitation will lead to a higher TMI threshold. 

To verify this, we calculate the TMI threshold for both the square and D-shaped fibers for FM-only excitation and equal mode excitation with increasing number of modes. A threshold increase factor (TIF) is defined by taking the ratio between the TMI thresholds for multimode excitation and FM-only excitation. The results are shown in Fig.~\ref{fig:Fig6}c, where we we have also plotted results for circular fiber as dotted curve for comparison. For both the square and D-shaped fibers, the TMI threshold increases linearly with the number of excited modes, similar to the circular fiber. This is in accordance with our reasoning based on the banded nature of the coupling matrix. The slope of the threshold scaling is highest for the D-shaped fiber leading to more than a $14\times$ higher TMI threshold, when all 82 modes are equally excited. The square shape leads to a slightly lower enhancement compared to the standard circular fiber. The superiority of the D-shaped fiber can be attributed to a lack of regular structure in the higher order modes.  As a result, unlike the more regular shapes, here no single temperature eigenmode has a particularly strong overlap with a given mode pair; instead, many temperature eigenmodes with different eigenvalues participate in the coupling, leading to a broader TMI gain spectrum with a lower peak value.
\end{appendices}

%


\newpage
\textbf{\large Supplementary Information}
 
 \renewcommand{\thesection}{S.\arabic{section}}
 \setcounter{section}{0} 

 \renewcommand{\theequation}{S.\arabic{equation}}
 \setcounter{equation}{0}

\renewcommand\thefigure{S.\arabic{figure}} 
  \setcounter{figure}{0}

	\section{Phase Matching}

A significant simplification of the coupled modal amplitude equations in the main text (Eq. [11]) describing the stimulated thermo-optical scattering is obtained by imposing the phase-matching condition, a standard procedure in nonlinear optics. Most terms on the right hand side (RHS) of Eq. [11] contributing to the scattering have complex phases and they vary in an oscillatory manner along the fiber axis. For sufficiently long fibers, the integrated scattering contribution of such terms is therefore negligible and can be ignored. Hence, a few special terms for which the relative phase is zero everywhere dominantly contribute to the scattering. These phase-matched terms satisfy the following condition:

\begin{equation}
    \beta_m - \beta_n + \beta_i - \beta_j=0 
    \label{Eq:PM},
\end{equation}

\noindent
where, $\beta_i$ is the propagation constant for the mode $i$.  In general, a number of solutions exist for Eq.~\ref{Eq:PM} giving rise to various phase matched terms. Each solution describes the contribution of the refractive index grating with the spatial frequency $q_{ij}=\beta_i -\beta_j$ in the total scattering.

	\subsection{Primary solution}

A main class of solutions to Eq.~\ref{Eq:PM} is given by $\beta_m=\beta_j$ and $\beta_n=\beta_i$. In the absence of exact degeneracies, this translates to $m=j$ and $n=i$. Physically, these terms represent the solutions where mode $n$ interferes with mode $m$ to create a spatially-varying heat source which gives rise to a refractive index grating with the right periodicity ${2\pi}/{(\beta_n-\beta_m)}$ to interact with light in mode $n$ and scatter power into mode $m$. Therefore, to describe the thermo-optical coupling that leads to TMI, these are the terms which we retain in Eq.~[16] in the main text. 
        \subsection{Additional solutions}

Another class of solutions to Eq.~\ref{Eq:PM} is given  by $\beta_i=\beta_j$ and $\beta_m=\beta_n$. These solutions represent the case where mode $i$ interacts with itself (typically with a Stokes shifted frequency) to create a uniform heat source ($q_{ij}\approx0$) which leads to a nearly uniform refractive index change resulting in a nonlinear phase modulation in all the other modes, $m$. Note that since the periodicity of the grating created in this case is very large (but finite due to the Stokes shift leading to slightly different propagation constants), approaching 
the typical length of the fiber amplifier, these terms do not lead to a transfer of power between various modes, and hence are ignored in Eq.~[16] in the main text. However, since these terms do lead to nonlinear self and cross phase modulation they can be important in describing the dynamics above the TMI threshold. It was shown in Refs.~\cite{zervas2021modal} and \cite{li2022threshold} that such terms along with other non-phase matched terms can actually lead to a behavior similar to modulation instability, even at zero Stokes frequency shift, at really high powers. But the threshold for such a process is typically much higher than that due to the phase-matched dynamic transfer of power leading to TMI in fibers with diameters smaller than 80 \textmu m~\cite{li2022threshold}.
        \subsection{Impact of degeneracies}

While retaining the phase-matched terms in Eq.~[16] in the main text, we also made an assumption that none of the fiber modes are exactly degenerate. This is justified because most high-power fiber amplifiers utilize step-index fibers and have imperfections and spatially inhomogeneous temperature distribution which typically break the exact circular geometry of the refractive index distribution in the fiber, leading to a lifting of mode degeneracies. However in the case when exact or nearly exact degeneracies are present, our formalism needs to take these into account, which can be achieved rather straightforwardly. We can replace Eq.~[17] in the main text with a slightly modified equation for the Stokes growth in each mode:

\begin{equation}
\begin{aligned}
    \frac{d B^\sigma_m(z,\Omega)}{dz} =&g \sum_{\sigma^\prime} (\frac{1}{2}\delta_{\sigma\sigma^\prime}\\& +i\chi_0\sum_{n}G^{\sigma\sigma\sigma^\prime\sigma^\prime}_{mnnm}(\Omega)A^{s*}_{n,\sigma}A^{s}_{n,\sigma^\prime}B^{\sigma^\prime}_m(z,\Omega)).
    \label{Eq:Nampgrowth}
\end{aligned}
\end{equation}

\noindent
Here we have introduced the index $\sigma$ for labelling modes within a degenerate subspace, in addition to the mode index $n$ that identifies various non-degenerate mode groups. For instance in perfect circular step-index fibers, the $\rm LPmn$ modes (for $m\neq0$) are two-fold degenerate, due to two solutions in the azimuthal direction (sine and cosine) for the same propagation constant, hence we have $\sigma=1,2$. The key consequence of modal degeneracy is that the equations for the growth of the Stokes power in the modes within a degenerate sub-group become coupled. The stokes growth rate  for each sub-group is given by the eigenvalue of the growth matrix whose elements are given by the weighted sum of the thermo-optical coupling (given by $G^{\sigma\sigma\sigma^\prime\sigma^\prime}_{mnnm}$) with weights equal to the signal amplitudes in various modes. Note that the equations for the Stokes growth in various non-degenerate mode groups are still uncoupled. This leads to an overall growth matrix for the Stokes amplitudes which has a block diagonal form with relatively small block sizes and can be easily diagonalized. In the case of sine/cosine degeneracy in LP modes, the size of the blocks is $2\times2$. For the graded-index fibers, the role of degeneracies can be more important since the size of degenerate sub-groups increases with order of the modes. 

An interesting consequence of the coupling of the Stokes growth equations for degenerate modes is that the Stokes growth rate typically becomes larger, as it is mainly determined by the maximum eigenvalue, which is increased by the presence of cross-couplings, while the smallest eigenvalue is suppressed. For 2-fold LP mode degeneracy, the TMI gain would be increased by a factor of 2. However, since such an effect would impact the TMI gain equally for both fundamental-mode excitation, where an LP01 mode couples with two degenerate LP11 modes, and for multimode excitations; hence it does not change the enhancement factor of TMI threshold upon multimode excitation. Note that the coupling to exactly degenerate HOM modes has never been considered in previous theories for TMI which assume single mode excitation, and some of these theories have shown good agreement to experimental results. A possible explanation for this is the likely breaking of exact degeneracies in realistic fiber amplifiers, as mentioned earlier

		\begin{figure*}[t!]
			\centering
			\includegraphics[width=0.5\textwidth]{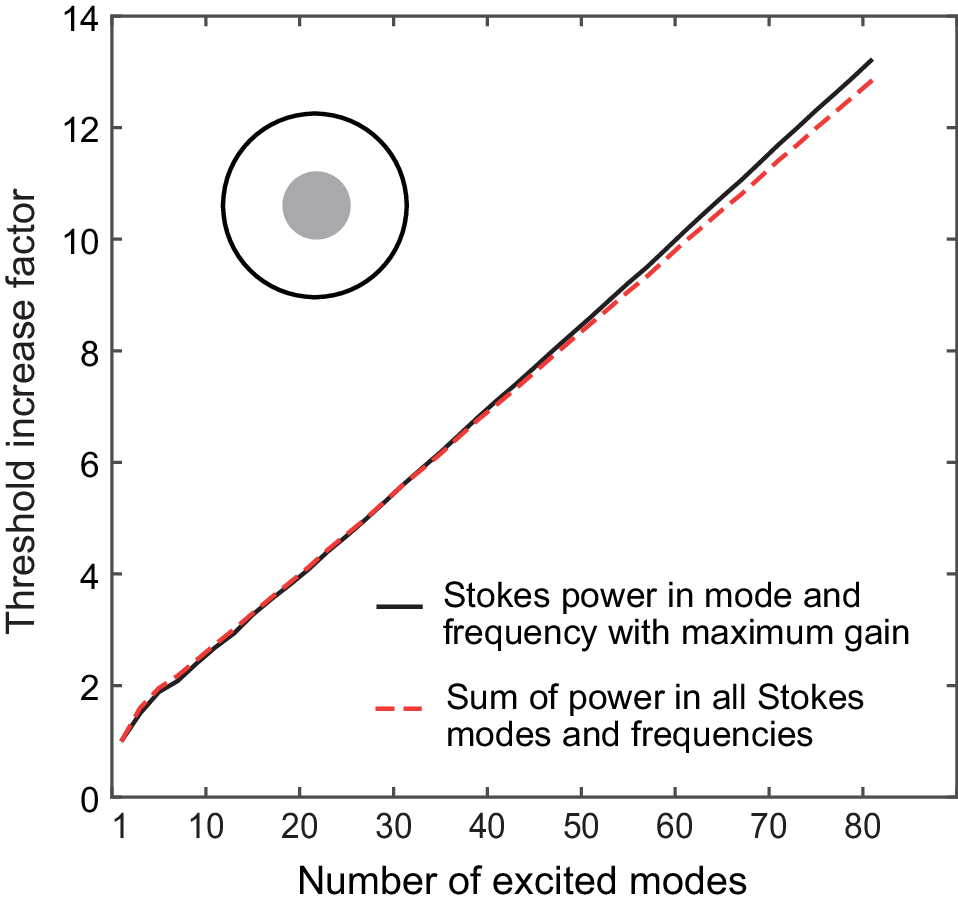}
		
		\caption{ \textbf{Scaling of TMI threshold with the number of equally excited modes in a multimode fiber} The approximate model (solid black curve) which only considers the Stokes power in the mode with maximum growth rate at peak gain frequency produces a very similar scaling as the full model (dashed red curve) which considers the total Stokes power. This is a result of the exponential nature of the Stokes power growth starting from a very small seed power. }        
		\label{fig:scalingSum}
            \end{figure*}

 \section{Multimode TMI threshold}

 In section II.C of the main text we have defined the multimode TMI threshold as the total output signal power at which the Stokes power reaches a significant fraction ($>1\%$) of the signal power. The noise power at the input is assumed to be seeded either by relative intensity noise due to signal linewidth or other experimental imperfections. In absence of these effects, the noise power has a lower limit given by the quantum noise and is approximately $10^{-16 } W$ per stokes mode in a KHz frequency window. We approximate the total noise power at the output by the power in the mode with maximum Stokes gain at the peak Stokes gain frequency. Although the noise power is present in multiple modes and frequencies, the approximation of considering only the mode and frequency with peak gain is still justified due to the exponential nature of the Stokes growth. Here we confirm the validity of this approximation both mathematically and with a concrete example. At the TMI threshold, the total noise power at the output is given by:

\begin{equation}
      P^N(L)=\sum_m \int_{-\infty}^{\infty} d\Omega \Tilde{P}_m(0,\Omega) e^{gL} e^{\bar{\chi}_m(\Omega)(P_{\rm th}-P(0))}. 
      \label{Eq:totalStokes}
\end{equation}

\noindent
Here $\bar{\chi}_m(\Omega)$ is the effective thermo-optical coupling for mode $m$, defined in Eq.~[21] in the main text. We assume a uniform density of input noise power ($\Tilde{P}_m(0,\Omega)$ = $\Tilde{P}^N(0)$) across a relevant frequency window in all fiber modes. Since the integrand in Eq.~\ref{Eq:totalStokes} is an exponential function we utilize saddle-point method to approximate the integral, leading to:

\begin{equation}
      P^N(L)\approx \sum_m  \Tilde{P}^N(0) e^{gL} \sqrt{\frac{2\pi}{\bar{\chi}_m^{\prime\prime}(\Omega_p)(P_{\rm th}-P(0))}} e^{\bar{\chi}_m(\Omega_p)(P_{\rm th}-P(0))}. 
\end{equation}
The total noise power depends exponentially on the total extracted signal power ($P_{\rm th}-P(0)$) and the effective thermo-optical coupling coefficient at the peak gain frequency $\Omega_p$. Here $\Tilde{P}^N(0)$ is the input noise power density with the units of [W/Hz]. The square root term in the prefactor depends on curvature (given by the second-order derivative) of the TMI gain curve and has units of [$\rm Hz$]. Further simplifications of the prefactor can be achieved by approximating $\bar{\chi}_m(\Omega)$ with a parabolic approximation around $\omega_p$ giving $\bar{\chi}_m^{\prime\prime}(\Omega_p)=\bar{\chi}_m(\Omega_p)/({\Delta \Omega})^2$, where $\Delta \Omega$ is the relevant frequency bandwidth for the thermo-optical coupling. This leads to the more familiar form for the Stokes power used in the main text:

\begin{equation}
      P^N(L)\approx \sum_m  P^N(0) \gamma e^{gL}  e^{\bar{\chi}_m(\Omega_p)(P_{\rm th}-P(0))}.
\end{equation}
Here, $P^N(0)=\Tilde{P}^N(0)\Delta\Omega$ is the total input noise power per mode in the relevant frequency window, and we have defined a unitless quantity $\gamma=\sqrt{{2\pi}/{\bar{\chi}_m(\Omega_p)(P_{th}-P(0))}}$, which is roughly fixed at the TMI threshold with a value on the order of unity. For our fiber, $\gamma \approx 0.5$. In any case, the prefactor in the above equation plays little role in determining the TMI threshold, which primarily depends on the argument of the exponential.

Further, the sum over the Stokes modes can be approximated by the largest contribution or a few largest contributions in the case that multiple Stokes modes have the same gain, leading directly to Eq.~[22] in the main text. This approximation is equivaluent to replacing the LogSumExp function (also called RealSoftMax) by the max function. It is known that LogSumExp is tightly bounded above and below by the max function up to small logarithmic corrections, which rigorously justifies our approximation~\cite{blanchard2021accurately}.

To validate the approximation of the total Stokes power by the Stokes power in the mode with maximum Stokes gain at the peak Stokes gain frequency, we computed the scaling of TMI threshold using both the exact expression for total Stokes power Eq.~\ref{Eq:totalStokes} and the approximate expression used in the main text Eq.~[22]. The results are shown in Fig.~\ref{fig:scalingSum}. It can be seen that the scaling of the TMI threshold computed by the two expressions is nearly identical. Approximating the Stokes power with the power in the mode with maximum Stokes gain at peak frequency is quite useful, as it allows a straightforward formula for TMI threshold in terms of a single axially-integrated thermo-optical coefficient, which linearly depends on the signal power in various modes.

\section{Gain Saturation}
\subsection{Gain Saturation parameters}
In Eq.~[27] in the main text, the following formula is used for $g_{mn}$, the gain in mode $m$ due to complex amplitude of field in mode $n$ in presence of gain saturation:
\begin{equation}
    g_{mn}(z) =  \frac{g_0}{2}\int d \vec{r}_\perp \frac{\psi^*_m (\vec{r}_\perp)\psi_n (\vec{r}_\perp)}{1+I_0 (\vec{r}_\perp,z)/I_{\rm sat} (z)}.
    \label{Eq:signalgain}
\end{equation}
Here we derive this formula along with the formulas for unsaturated gain coefficient $g_0$ and signal saturation intensity $I_{\rm sat}$ from the steady-state solution to the rate equations for Yb doped gain medium. In the steady state, the fraction of the population in the upper lasing level $n_u$ is given by~\cite{paschotta1997ytterbium,smith2011mode}:
\begin{equation}
    n_{\rm u} = \frac{\frac{P_{\rm p}}{A_{\rm cl}\hbar\omega_{\rm p}}\sigma_{\rm p}^{\rm a}+\frac{I_0}{\hbar\omega_{\rm s}}\sigma_{\rm s}^{\rm a}}{\frac{P_{\rm p}}{A_{\rm cl}\hbar\omega_{\rm p}}(\sigma_{\rm p}^{\rm e}+\sigma_{\rm p}^{\rm a})+\frac{I_0}{\hbar\omega_{\rm s}}(\sigma_{\rm s}^{\rm e}+\sigma_{\rm s}^{\rm a})+\frac{1}{\tau}}.
    \label{Eq:nu}
\end{equation}
Here, $\omega_{\rm s}$ is the signal frequency, $\omega_{\rm p}$ is the pump frequency, $\sigma_{\rm s}^{\rm e}$ and $\sigma_{\rm s}^{\rm a}$ are the emission and absorption cross-sections of the signal, $\sigma_{\rm p}^{\rm e}$ and $\sigma_{\rm p}^{\rm a}$ are the emission and absorption cross sections of the pump, $P_{\rm p}$ is the local pump power, and $I_0$ is the local signal intensity. This formula can be rewritten as follows:

\begin{equation}
    n_{\rm u} = \frac{\frac{P_{\rm p}}{P_{\rm p}^{\rm sat}}\frac{\sigma_{\rm p}^{\rm a}}{\sigma_{\rm p}^{\rm e}+\sigma_{\rm p}^{\rm a}}+\frac{I_0}{I^0_{\rm sat}}\frac{\sigma_{\rm s}^{\rm a}}{\sigma_{\rm s}^{\rm e}+\sigma_{\rm s}^{\rm a}}}{1+\frac{P_{\rm p}}{P_{\rm p}^{\rm sat}}+\frac{I_0}{I^0_{\rm sat}}},
    \label{Eq:nusimp}
\end{equation}
\noindent
where $P^{\rm sat}_{\rm p}$ is the pump saturation power and $I^0_{\rm sat}$ is signal saturation intensity at zero pump power and these are defined as follows:
\begin{equation}
    P^{\rm sat}_{\rm p} = \frac{A_{\rm cl}\hbar\omega_{\rm p}}{\tau(\sigma_{\rm p}^{\rm e}+\sigma_{\rm p}^{\rm a})},\:\:\:\:I^0_{\rm sat}= \frac{\hbar\omega_{\rm s}}{\tau(\sigma_{\rm s}^{\rm e}+\sigma_{\rm s}^{\rm a})}.
\end{equation}
\noindent
Since $\sigma^a_s \ll \sigma^e_s$, the second term in the numerator in Eq.~\ref{Eq:nusimp} can be ignored, especially when significant pump power is present in the fiber. Next, we divide both the numerator and denominator in Eq.~\ref{Eq:nusimp} by $1+\frac{P_p}{P_p^{\rm sat}}$ to obtain a familiar form for the upper level population:
\begin{equation}
    n_{\rm u} \approx \frac{n_u^0}{1+I_0/I_{\rm sat}},
    \label{Eq:nu0}
\end{equation}
where $n_{\rm u}^0$ is the maximum upper level population (occurring when the signal intensity is zero) and $I_{\rm sat}$ is the pump-dependent signal saturation intensity and these are given by:
\begin{equation}
    n_u^0 \approx \frac{P_p}{P_p+P_p^{\rm sat}}\frac{\sigma_{\rm p}^{\rm a}}{\sigma_{\rm p}^{\rm e}+\sigma_{\rm p}^{\rm a}},\:\:\:\:I_{\rm sat}= I^0_{\rm sat}(1+P_p/P_p^{\rm sat}).
     \label{Eq:nuapprox}
\end{equation}
Note that the peak fractional upper level population, which defines the unsaturated gain coefficient $g_0$ depends on both the pump power, and the properties of the gain medium. At low pump powers, $n_u^0$ is directly proportional to the pump power but for $P_p \gg P_p^{\rm sat}$, $n_u^0$ is roughly constant.   The formula for signal saturation intensity also shows a dependence on the pump power along with various properties of the gain medium. At small pump power ($P_{\rm p} \ll P_{\rm p}^{\rm sat}$), $I_{\rm sat}$ is roughly constant and is equal to $I_{\rm sat}^0$. However in high-power fiber amplifiers typically the starting pump power is significantly higher than $P_{\rm p}^{\rm sat}$, resulting in $I_{\rm sat}$ varying linearly with the pump power. Hence pump depletion and gain saturation are interconnected in such amplifiers.  The upper level population can be used to calculate signal gain as follows:
\begin{equation}
\begin{aligned}
    g_{mn}(z) &=  \frac{1}{2}N_{\rm Yb}\int d\vec{r}_\perp \psi_m^* \psi_n(\sigma_s^e n_{\rm u}(\vec{r}_\perp,z) - \sigma_s^a n_{\rm l}(\vec{r}_\perp,z)) \\& \approx \frac{1}{2} N_{\rm Yb} \sigma_s^e\int d\vec{r}_\perp \psi_m^* \psi_n n_{\rm u}(\vec{r}_\perp,z). 
    \label{Eq:gmnint}
\end{aligned}
\end{equation} 

\noindent
Here $N_{\rm Yb}$ is the density of doped Ytterbium ions. The term involving the lower level population $n_{\rm l}$ is ignored since $\sigma_{\rm s}^{\rm a} \ll \sigma_{\rm s}^{\rm e}$. Substituting the expression for $n_u$ from Eq.~\ref{Eq:nu0} into Eq.~\ref{Eq:gmnint}, we obtain the formula for $g_{mn}$ given in Eq.~\ref{Eq:signalgain}, and the unsaturated gain coefficient $g_0$ is equal to:
\begin{equation}
    g_0 = N_{\rm Yb}\sigma^{\rm e}_{\rm s} n_{\rm u}^0 \approx N_{\rm Yb} \sigma^{\rm e}_{\rm s} \frac{P_{\rm p}}{P_{\rm p}+P_{\rm p}^{\rm sat}}\frac{\sigma_{\rm p}^{\rm a}}{\sigma_{\rm p}^{\rm e}+\sigma_{\rm p}^{\rm a}}
\end{equation}

\subsection{Dynamic heat load saturation}

In Eq.~[32] in the main text, we provide the following formula for the saturated dynamic heat load:

\begin{equation}
    \Tilde{Q}(\vec{r},t) \approx  \frac{g_0 q_{\rm D} \Tilde{I}(\vec{r},t)}{{(1+I_0(\vec{r})/I_{\rm sat})}^2}.
    \label{Eq:satdynamicheatSrc}
\end{equation}
\noindent
The quadratic power of the saturation term in the denominator was first pointed out by Hansen et al.~\cite{hansen2014impact} and is responsible for the increased TMI threshold due to the gain saturation. Here we provide a detailed derivation for this formula. The total gain-saturated heat load is given by~\cite{hansen2014impact}:

\begin{equation}
    Q(\vec{r},t)= g_{\rm s} q_{\rm D} I(\vec{r},t)=\frac{g_0 q_{\rm D} I(\vec{r},t)}{1+I(\vec{r},t)/I_{\rm sat}},
    \label{Eq:satheatSrc}
\end{equation}
where, $g_{\rm s}$ and $g_0$ are the saturated and unsaturated gain coefficients and $q_{\rm D}$ is the quantum defect. At any point, the signal intensity can be written as a sum of a static contribution $I_0(\vec{r})$ and a dynamic contribution $\Tilde{I}(\vec{r},t)$, i.e., $I(\vec{r},t)=I_0(\vec{r})+\Tilde{I}(\vec{r},t)$.
Substituting this in Eq.~\ref{Eq:satheatSrc} and rearranging the denominator we obtain:
\begin{equation}
\begin{aligned}
    Q(\vec{r},t)&=\frac{g_0 q_{\rm D}(I_0+\Tilde{I})}{(1+I_0/I_{\rm sat})(1+\Tilde{I}/(I_0+I_{\rm sat}))}\\& =\frac{g_0 q_{\rm D}}{(1+I_0/I_{\rm sat})}(I_0+\Tilde{I}){(1+\Tilde{I}/(I_0+I_{\rm sat}))}^{-1}.
    \label{Eq:satheatreorder}
\end{aligned}
\end{equation}
Here we have suppressed the arguments of all the functions for compactness. Below the TMI threshold the dynamic signal intensity $\Tilde{I}$ is much smaller than $I_0$ and $I_{\rm sat
}$, therefore we can Taylor-expand the last term in the numerator:
\begin{equation}
    Q(\vec{r},t)=\frac{g_0 q_{\rm D}}{(1+I_0/I_{\rm sat})}(I_0+\Tilde{I}){(1-\Tilde{I}/(I_0+I_{\rm sat})+\mathcal{O}(\Tilde{I}^2))}.
    \label{Eq:satheattaylor}
\end{equation}
Multiplying the terms in the numerator and ignoring the terms which are $\mathcal{O}(\Tilde{I}^2)$ and higher, we get:
\begin{equation}
\begin{aligned}
    Q(\vec{r},t)&\approx\frac{g_0 q_{\rm D}}{(1+I_0/I_{\rm sat})}{(I_0+\Tilde{I}(1-I_0/(I_0+I_{\rm sat})))}\\& =\frac{g_0 q_{\rm D}I_0}{(1+I_0/I_{\rm sat})}+\frac{g_0 q_{\rm D}\Tilde{I}}{{(1+I_0/I_{\rm sat})}^{2}}
    \label{Eq:satheatfinal}
\end{aligned}
\end{equation}
\noindent
The first term on the right hand side of the above equation describes the static heat load, and the last term  gives the dynamic heat load, which has an exact form used in the main text and shown in Eq.~\ref{Eq:satdynamicheatSrc}.

\begin{figure*}[t!]
    \centering
    \includegraphics[width=\textwidth]{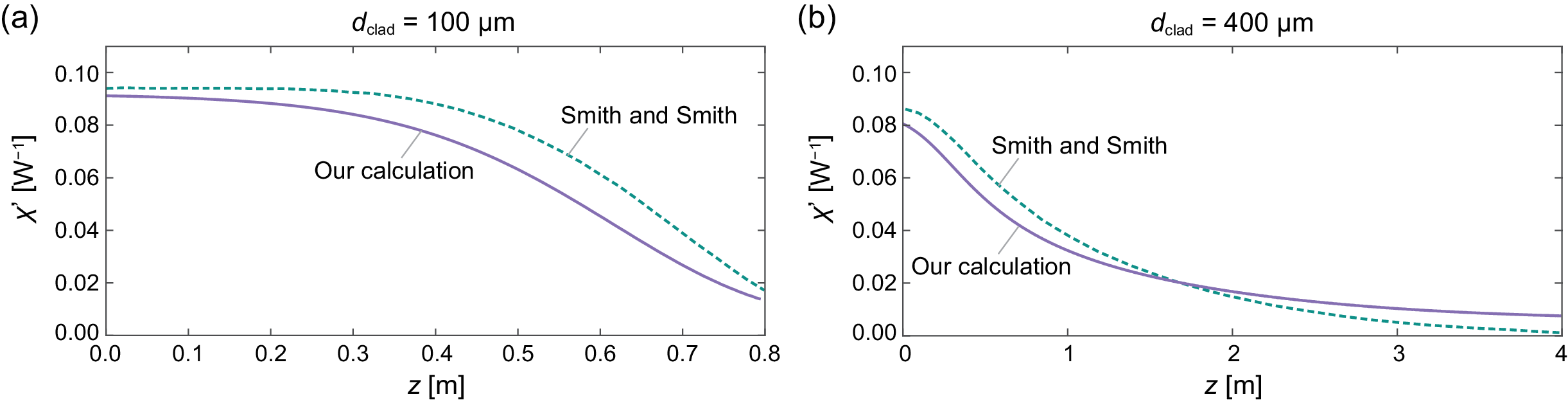}
    \caption{Saturated thermo-optical coupling $\chi\prime$ between the fundamental mode and first higher order mode (HOM) computed using our TMI model (solid curves) and compared to detailed simulations in Ref.~\cite{Smith2013increasing} (dotted curves). Fibers with two different cladding radii (with different gain saturation amount) are considered. Our results are in close agreement with those in Ref.~\cite{Smith2013increasing}.}
    \label{fig:Fig3}
\end{figure*}

\begin{figure*}[t!]
    \centering
    \includegraphics[width=\textwidth]{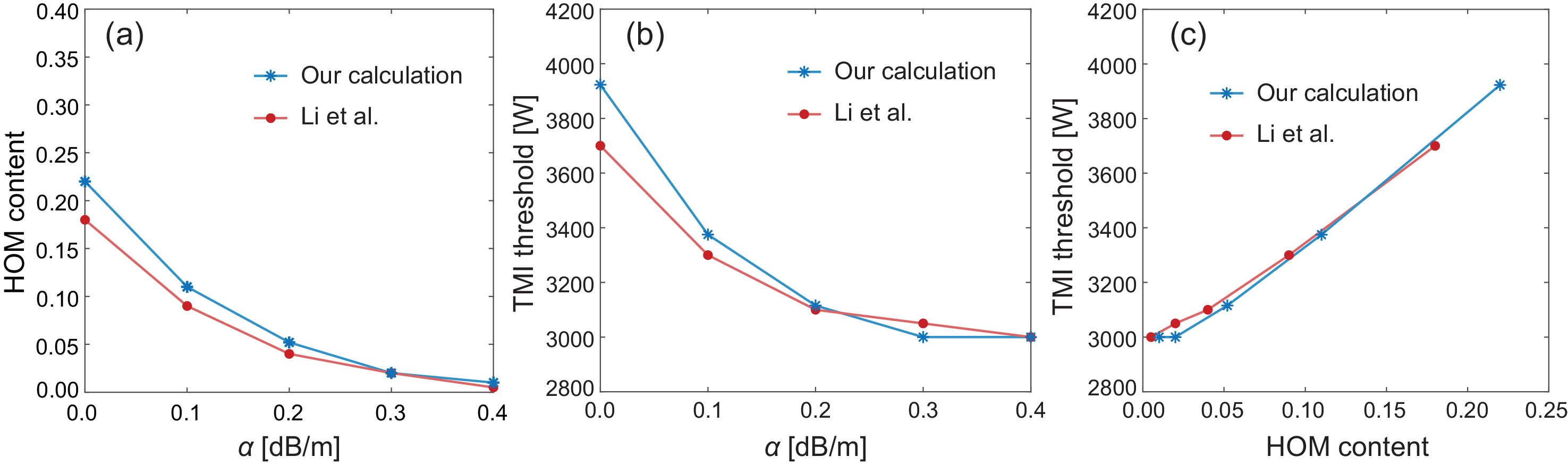}
    \caption{Comparison between results of our TMI model and Ref.~\cite{li2023mitigation} by Li et al. for TMI threshold upon variable two-mode excitation. (a) As the HOM loss $\alpha$ is increased, the relative HOM content decreases. (b) TMI threshold increases with decreasing $\alpha$ as the excitation becomes more multimode (c) TMI threshold scales roughly linearly with the HOM content. Overall the results from Li et al. are in close agreement with our TMI model.}
    \label{fig:Fig4}
\end{figure*}

\section{Comparison with previous studies}

In this section, we perform feasible comparisons/validations to experiment and previous theories as described below. 
\newline

\noindent\textbf{(1) Experimental results - Single mode excitation: Ref.~\cite{Li2017experimental}.}

Using our model, we calculated the TMI threshold for the co-pumped 50/400 fiber discussed in Ref.~\cite{Li2017experimental} operating at 1064 nm signal wavelength and obtained a value of 680 W which closely matches with the experimentally obtained value of 658 W shown in Table II in Ref.~\cite{Li2017experimental}. For the noise seed power, we utilize the value of relative intensity noise (RIN) provided in Table I ($R_N = 10^{-10}\rm \:Hz^{-1}$) in Ref.~\cite{Li2017experimental}. The axially-integrated TMI gain to reach the TMI threshold is roughly 13.8.   Since there are few existing experimental or detailed numerical studies of TMI for highly multimode excitations prior to our work we have validated our model against existing studies on TMI for single mode and two-mode excitations. Two examples of detailed validation of our model are presented below:
\newline

\noindent\textbf{(2) Single mode excitation: Ref.~\cite{Smith2013increasing}.}

One of the earliest studies on the impact of gain saturation on the TMI threshold was published by Smith et al.\cite{Smith2013increasing} Using detailed numerical simulations, they studied the thermo-optical coupling $\chi\prime$ between the fundamental mode and the first higher order mode in a fiber with core diameter of $50 \: \mu m$. It was shown that in the presence of gain saturation the effective thermo-optical coupling coefficient is lowered due to the reduction of dynamic heat load. The results from Fig.5 in Ref.~\cite{Smith2013increasing} are reproduced here as dotted curves in Fig.S2. We use our model to calculate the effective thermo-optical coupling for a co-pumped fiber with same parameters as used in Ref.~\cite{Smith2013increasing}. Our results (solid curves in Fig.S2) are in excellent agreement with the detailed simulation results in Ref.~\cite{Smith2013increasing}. When a fiber with larger cladding (pump core) diameter is used (Fig. S2b) the amount of gain saturation is higher as the pump intensity in the gain core is lowered. This leads to an increased reduction in the thermo-optical coupling and a higher TMI threshold. We also calculate the TMI threshold for the both the fibers ($d_{\rm clad} = 100 \mu m$ and $d_{\rm clad} = 400 \mu m$) using our model and the values (488 W and 1101 W) match closely with the values (495 W and 1100 W) in Ref.~\cite{Smith2013increasing}. For noise seeding, we consider quantum noise ($P_N=10^{-16} W$) to match the parameters given in Table I in Ref.~\cite{Smith2013increasing}. The axially-integrated TMI gain to reach the TMI threshold is roughly 34.5.
\newline

\noindent\textbf{(3) Two mode excitation: Ref.~\cite{li2023mitigation}}

A recent study which utilized detailed simulations to calculate the TMI threshold upon two-mode excitation was presented in Li et al. in Ref.~\cite{li2023mitigation}. The gain fiber was operated at a signal wavelength of 1080 nm and a pump wavelength of 1018 nm. By varying the higher order mode (HOM) loss coefficient $\alpha$ they were able to achieve a two-mode excitation with variable HOM content. Fig. S3a shows the HOM content as $\alpha$ is increased calculated using our model for the co-pumped case and that presented in Fig.S3b in Ref.~\cite{li2023mitigation}. As $\alpha$ is increased HOM content decreases and the values calculated with our model match reasonably with Ref.~\cite{li2023mitigation}. For each value of $\alpha$ the TMI threshold is obtained with our model, and it agrees with the prior result. 
For noise seeding, we consider quantum noise ($P_N\approx4
\times10^{-16} W$) to match the parameters given in Ref.~\cite{li2023mitigation}. The axially-integrated TMI gain to reach the TMI threshold is roughly 34.5.
Figure S3b shows that for smaller $\alpha$ (higher HOM content) the TMI threshold is higher as the effective number of modes are higher. This is explicitly shown in Fig.S3c where we plot the TMI threshold as a function of HOM content which shows linear scaling with the HOM content and our results match closely with those in Ref.~\cite{li2023mitigation}. Note that the linear increase is expected to work only until the HOM content is less than  0.5 in the two-mode case after which the effective number of modes would start to decrease.

\end{document}